# Time-dependent modelling of oscillatory instability of three-dimensional natural convection of air in a laterally heated cubic box

**Alexander Yu. Gelfgat**

School of Mechanical Engineering, Faculty of Engineering, Tel-Aviv University, Ramat Aviv, Tel-Aviv, Israel, 69978, gelfgat@tau.ac.il

**Abstract**

Transition from steady to oscillatory buoyancy convection of air in a laterally heated cubic box is studied numerically by straight-forward time integration of Boussinesq equations using a series of gradually refined finite volume grids. Horizontal and spanwise cube boundaries are assumed to be either perfectly thermally conducting or perfectly thermally insulated, which results in four different sets of thermal boundary conditions. Critical Grashof numbers are obtained by interpolation of numerically extracted growth/decay rates of oscillations amplitude to zero. Slightly supercritical flow regimes are described by time-averaged flows, snapshots, and spatial distribution of oscillations amplitude. Possible similarities and dissimilarities with two-dimensional instabilities in laterally heated square cavities are discussed. Break of symmetries and sub- or super-critical character of bifurcations are examined. Three consequent transitions from steady to oscillatory regime, from oscillatory to steady, and finally to oscillatory flow, are found in the case of perfectly insulated horizontal and spanwise boundaries. Arguments for grid and time step independence of the results are given.

**Key words**: natural convection, instability, direct numerical simulation

# 1. Introduction

This study revisits a well-known problem of oscillatory instability of buoyancy convection in laterally heated cubic cavity. More than a decade ago a significant effort was invested in the study of instability of convection in laterally heated two-dimensional cavities. Cavities of different aspect ratios and fluids with different Prandtl numbers were taken into consideration. For more details and relevant literature the reader is referred to the book [1]. It is well-known nowadays that the primary instability of steady buoyancy convection flows in laterally heated cavities sets in via a Hopf bifurcation and results in an oscillatory flow regime. It is known also that flows in horizontally or vertically elongated cavities exhibit multiplicity of asymptotically stable steady [2] and oscillatory [3] states. Mathematical and numerical approaches for path-following and stability analysis, especially with respect to fluid dynamics problems, are being developed continuously, see, e.g., reviews [4-6] . While several decades ago, study of stability of a numerically calculated flow was a challenging state-of-the-art computational task, with the development of computational tools and computer power, study of stability or state multiplicity in a *two-dimensional* flow model became feasible and can be performed by means of several numerical tools [6]. Same tools can be applied when stability of two-dimensional base flow with respect to three-dimensional perturbations periodic in one spatial direction is studied [3,7]. However, it remains challenging when basic flow is three-dimensional, so that no simplifications such as spatial periodicity assumptions can be made. The latter was defined as Tri-Global linear stability problem in [4,5].

Among variety of theoretical and applied TriGlobal problems, two widely accepted benchmark configurations are usually considered. The first one is flow in a lid-driven cubic cavity, for which the primary instability was recently studied, and the results were verified by independent numerical approaches [8-11], as well as validated in an experiment [12]. The second one is linear instability of buoyancy convection of air (assuming the Prandtl number *Pr*=0.71) in a laterally heated cube. The later problem was studied experimentally in [13]. Early numerical studies considered perfectly thermally conducting horizontal boundaries of the cube [14,15], like in above experiment. Later studies focused on a more challenging problem with perfectly thermally insulated horizontal boundaries [20-19]. The spanwise boundaries were taken as perfectly thermally insulated in all above studies, so that effect of these boundary conditions was never studied. All above studies performed straight-forward time integration of the three-dimensional Boussinesq equations.

As is known from well-established numerical solutions obtained for instability of convective air flow in a two-dimensional laterally heated square cavities, which is a TwoGlobal problem according to [4,5], the critical Grashof number for perfectly thermally conducting horizontal boundaries is of the order $10^6$, while for perfectly thermally insulated borders it becomes larger than $10^8$ [2,20,22]. This makes the problem with perfectly thermally conducting boundaries computationally easier, so that 3D results obtained in [14,15] compare rather well with the experimental measurements of [13]. At the same time, the critical parameters obtained for the primary stability of 3D convection in a cube with perfectly thermally insulated horizontal boundaries [16-19] exhibit a noticeable scatter. The main reason for the scatter can be insufficient numerical convergence. Most of the cited works report grid convergence of some parameters of steady state flows, however none of them reported how the critical Grashof (Rayleigh) number varies with grid refinement. Grid convergence of the oscillations frequency of a computed slightly supercritical state was reported only in [20].

It already became a common knowledge that a numerically accurate solution of a stability problem usually requires finer resolution than computation of a steady flow only. The finer resolution is needed because, along with the steady state flow, one has to calculate the most unstable eigenmode, which can be more difficult, than computation of the flow itself. A classic example of the latter is 2D configuration with thermally insulated horizontal boundaries, in which instability sets in inside very thin boundary layers adjacent to the hot and cold walls [7,21-23]. More results on grid convergence in 2D benchmark problems can be found in [22]. When instability is simulated by a straight-forward time integration, similar question arise regarding the time-step and time-integration-method independence, but, as a rule, are not addressed.

The primary goal of this study was to find critical Grashof numbers at which the steady state flows lose stability and become oscillatory. Performing computations on the gradually refined grids, having $100^3$, $150^3$, $200^3$, and $256^3$ finite volumes, we intended to arrive to converged values of the critical Grashof numbers, as well as to obtain numerically converged critical frequencies of appearing oscillations. Together with that we obtain new results on the symmetry preservation or breaking in slightly supercritical flows. Also, we make some arguments on the sub- and super-criticality of the bifurcations observed. The most unexpected and never reported was a cascade of three consequent steady-oscillatory transitions observed in the AA – AA case.

In the present study the instability onset and slightly supercritical oscillatory regimes of air convection in a cubic laterally heated box are studied by a straight-forward time integration of the Boussinesq equations. To illustrate three-dimensional velocity patterns we use method of quasi-two-dimensional divergence free projections recently proposed in [24,25]. All the boundaries are considered as no-slip. Two opposite vertical boundaries are assumed to be isothermal, while the upper and lower horizontal boundaries are considered either perfectly thermally conducting, or perfectly thermally insulated. Also, the second pair of the opposite vertical boundaries, which we call spanwise boundaries, is assumed to be perfectly conducting or perfectly insulated. Thus, four sets of different thermal boundary conditions are taken into consideration. This extends the benchmark data for future 3D stability studies, and offers to consider two "easier" cases with the perfectly conducting horizontal boundaries, as well as two "difficult" cases defining the horizontal boundaries as thermally insulated. Together with the results on lid-driven cubic cavity [8-11], these form a quite representative set of three-dimensional stability benchmark problems.

Considering perfectly thermally conducting horizontal boundaries we have found that steady – oscillatory transition there takes place similarly to the corresponding 2D case. The critical values of the Grashof number, as well as oscillation frequencies, remain close for two sets of the temperature spanwise boundary conditions. At the same time, the flow reflection symmetry sustains after the transition in the case of conducting spanwise boundaries, while all the symmetries break if these boundaries are thermally insulated. Also, development of the supercritical regimes at larger Grashof numbers appears to be different.

The case of perfectly conducting spanwise and perfectly insulated horizontal boundaries was never considered before and the corresponding results are reported here for the first time. The oscillatory instability in this case appears at $Gr \approx 1.25 \cdot 10^8$ as low-frequency and low-amplitude oscillations. No similarities with the corresponding two-dimensional case are observed. The observed transition breaks rotation and 2D reflection symmetries, but remains symmetric with respect to their superposition, which is considered as a rare and exceptional case.

When the spanwise boundaries are perfectly insulated a sequence of three bifurcations is observed. The first one appears at $Gr \approx 4.6 \cdot 10^7$, with a considerably smaller frequency and amplitude than in the previous CC – CC and CC – AA cases. This bifurcation was already observed in the previous studies [16,17,19], however their results for the critical Grashof number and the oscillations frequency agree only in the first significant digit. The transition breaks all the symmetries of the base flow and exhibits some partial similarities with the analogous 2D case. The next oscillatory – steady transition takes place when the Grashof number exceeds the value of $Gr\approx7.1\cdot10^7$, after which stability and symmetries of the steady flow are restored. The flow remains stable until $Gr\approx2.8\cdot10^8$, where the third steady-oscillatory transition takes place. This transition preserves the flow symmetries, but resulting oscillatory regime becomes unstable already at $Gr\approx3\cdot10^8$, beyond which the flow is multi-frequent and non-symmetric. There is some evidence showing that symmetric and non-symmetric oscillatory flows remain stable at the same Grashof number. The second and the third transition, as well as existence of multiple stable oscillatory flows, are reported here for the first time.

We are aware that pseudo-spectral and collocation methods may yield better accuracy for model problems in simple geometries, as those considered here. At the same time we are interested in the three-dimensional modeling and stability analysis of more complicated flows developing inside and around bodies of arbitrary shape. Besides a wide set of curved-boundary-fitted numerical methods, the latter can be achieved by combining of the fixed grid finite volume approach used below with the immersed boundary technique, as it was done recently in [26]. Our most representative $256^3$ nodes grid is slightly finer than those used in the recent studies of the primary instability in the lid-driven cavity [8-11], where the finest grid had only $200^3$ nodes.

In the following we give a brief formulation of the problem (Section 2) and describe two independent time-integration techniques applied (Section 3). In Section 4 we present examples of test calculations and numerical verification of the results. In Section 5 we explain the method of visualization of three-dimensional velocity field [24,25] and applied here. Results on transition to unsteadiness with the necessary discussion are presented in Section 6. Conclusions followed by a short discussion are summarized in Section 7.

## 2. Formulation of the problem

We consider natural convection of an incompressible fluid in a cubic cavity, whose opposite sidewalls are kept at constant and different temperatures $T_{hot}$ and $T_{cold}$. The flow is described by a set of Boussinesq equations that are rendered dimensionless taking the cube side length $H$ as a characteristic scale, and $H^2/\nu$, $\nu/H$, and $\rho \nu^2/H^2$ as scales of the time $t$, the velocity $\boldsymbol{v} = (u,v,w)$ and the pressure $p$, respectively. Here $\nu$ is the fluid kinematic viscosity and $\rho$ is the density. The temperature is rescaled to a dimensionless function using the relation $T \to (T - T_{cold})/(T_{hot} - T_{cold})$. Additionally, the dimensionless time, velocity and pressure are scaled, respectively by $Gr^{-1/2}$, $Gr^{1/2}$, and $Gr$, where $Gr = g\beta(T_{hot} - T_{cold})H^3/\nu^2$ is the Grashof number, $g$ is the gravity acceleration and $\beta$ is the thermal expansion coefficient. The resulting system of momentum, energy and continuity equations is defined in a cube $0 \leq x, y, z \leq 1$ and reads

$$\frac{\partial T}{\partial t} + (\boldsymbol{v} \cdot \nabla)T = \frac{1}{PrGr^{1/2}} \Delta T \tag{1}$$

$$\frac{\partial \boldsymbol{v}}{\partial t} + (\boldsymbol{v} \cdot \nabla)\boldsymbol{v} = -\nabla p + \frac{1}{Gr^{1/2}} \Delta \boldsymbol{v} + T\boldsymbol{e}_z \tag{2}$$

$$\nabla \cdot \boldsymbol{v} = 0. \tag{3}$$

Here $Pr = \nu/\alpha$ is the Prandtl number, and $\alpha$ is the thermal expansion coefficient. All the boundaries are assumed to be no-slip. Two vertical boundaries at $x = 0,1$ are kept isothermal, so that

$$T(x = 0, y, z) = 1, \quad T(x = 1, y, z) = 0. \tag{4}$$

Other boundaries are assumed to be either perfectly thermally conducting or perfectly thermally insulated. Following notations introduced in [27] we denote a pair of perfectly insulated (adiabatic) boundaries as AA, and a pair of perfectly conducting boundaries as CC. The four sets of thermal boundary conditions considered below are

CC – CC: $\quad T(x, y = 0, z) = T(x, y = 1, z) = T(x, y, z = 0) = T(x, y, z = 1) = 1 - x \tag{5}$

CC – AA: $\quad T(x, y = 0, z) = T(x, y = 1, z) = 1 - x, \quad \left(\frac{\partial T}{\partial z}\right)_{z=0} = \left(\frac{\partial T}{\partial z}\right)_{z=1} = 0 \tag{6}$

AA – CC: $\quad \left(\frac{\partial T}{\partial y}\right)_{y=0} = \left(\frac{\partial T}{\partial y}\right)_{y=1} = 0, \quad T(x, y, z = 0) = T(x, y, z = 1) = 1 - x \tag{7}$

AA – AA: $\quad \left(\frac{\partial T}{\partial y}\right)_{y=0} = \left(\frac{\partial T}{\partial y}\right)_{y=1} = 0, \quad \left(\frac{\partial T}{\partial z}\right)_{z=0} = \left(\frac{\partial T}{\partial z}\right)_{z=1} = 0. \tag{8}$

As is noted in [17], the problem has three symmetries: (i) reflection symmetry with respect to the midplane y=0.5, $\{u, v, w, \theta\}(x, y, z) = \{u, -v, w, \theta\}(x, 1 - y, z)$, (ii) 2D rotational symmetry with respect to rotation in $180°$ around the line x=z=0.5, $\{u, v, w, \theta\}(x, y, z) = -\{u, -v, w, \theta\}(1 - x, y, 1 - z)$ and (iii) 3D centro-symmetry $\{u, v, w, \theta\}(x, y, z) = -\{u, v, w, \theta\}(1 - x, 1 - y, 1 - z)$. Here $\theta = T - (1 - x)$. These symmetries are characteristic for steady state flows, however can be broken by instability, so that supercritical oscillatory flows can maintain only one of them or be fully non-symmetric. Clearly, the centro-symmetry (iii) is a superposition of the two other symmetries, so that seemingly it cannot be preserved if one of the symmetries (i) or (ii) is broken. However there exists an exceptional case. Assume that both symmetries (i) and (ii) are broken. Then the coordinate transformations in (i) and (ii) can be presented as a sum of symmetric and antisymmetric parts. If the antisymmetric parts of (i) and (ii) cancel each other, the 3D centro-symmetry (iii) persists. We follow the above symmetries by examination of the maximal absolute difference of the values of $\theta$ in the points of symmetry. In particular, we observe the above exceptional case in the AA – CC configuration (see below).

To compare results with the corresponding two-dimensional model of buoyancy convection in a square cavity, in the following we call areas adjacent to the cube edges (0,*y*,0), (0,*y*,1), (1,*y*,0) and (1,*y*,1) as lower left, upper left, lower right, and upper right corners, respectively.

The primary goal of the $Gr^{1/2}$ scaling is to make the velocity values of the order of unity, which allows for more accurate calculations. The form of equations (1) and (2) shows also that $Gr^{1/2}$ yields an estimate of the Reynolds number, as is known for natural convection flows (see, e.g., [28]). As was shown in [29] the same scale is applied to the vertical velocity in the boundary layers developing near vertical heated walls. Also, as is shown in [30], in the CC – AA case the Brunt-Väisälä frequency is estimated also by $\sqrt{g\beta(T_{hot} - T_{cold})H}/H$, so that its dimensionless value obtained using the viscous time scale $H^2/\nu$ is $\sqrt{g\beta(T_{hot} - T_{cold})H^3/\nu^2} = Gr^{1/2}$. Thus, the dimensionless time is scaled additionally by the inverse Brunt-Väisälä frequency.

### 3. Numerical method

The problem is solved using standard finite volume discretization on staggered grids. The resulting schemes yield the second order approximation in space, and conserve mass, momentum and internal energy. The scheme conservative properties are considered as a necessary condition for correct numerical identification of the instability threshold. The details are given in [31]. The size of stretched finite volume grid varies from $100^3$ to $256^3$. The stretching is the same as in [22].

For the time integration we apply the same second-order backward scheme and two independent pressure/velocity segregated or coupled approaches. The first one is a fractional step method with the second order discretization of time derivative. This scheme is similar to one applied in [14,15] and is briefly described below. Denoting the time step by $\delta t$ and by the superscript $(n)$ the function values at $t=n\delta t$, we perform the time integration as

$$\frac{1}{2\delta t}\left(3T^{(n+1)} - 4T^{(n)} + T^{(n-1)}\right) + \left(\boldsymbol{v}^{(n)} \cdot \nabla\right)T^{(n)} = \frac{1}{PrGr^{1/2}}\Delta T^{(n+1)} \tag{9}$$

$$\frac{1}{2\delta t}\left(3\boldsymbol{v}^{(n+1/2)} - 4\boldsymbol{v}^{(n)} + \boldsymbol{v}^{(n-1)}\right) + \left(\boldsymbol{v}^{(n)} \cdot \nabla\right)\boldsymbol{v}^{(n)} =$$
$$-\nabla p^{(n)} + \frac{1}{Gr^{1/2}}\Delta \boldsymbol{v}^{(n+1/2)} + T^{(n+1)}\boldsymbol{e}_z \tag{10}$$

$$\Delta(\delta p) = \frac{3}{2\delta t} div\, \boldsymbol{v}^{(n+1/2)} \tag{11}$$

$$p^{(n+1)} = p^{(n)} + \delta p, \quad \boldsymbol{v}^{(n+1)} = \boldsymbol{v}^{(n+1/2)} - \frac{2\delta t}{3} grad(\delta p). \tag{12}$$

First, we compute $T^{(n+1)}$ using Eq. (9), and $\boldsymbol{v}^{(n+1/2)}$ from Eq. (10). Then we calculate the pressure correction from the Poisson equation (11), which is supplied by the Neumann boundary conditions for the pressure. Finally, pressure and velocity are updated according to Eqs. (12). Note that 3/2 coefficient in Eq. (11) follows from the second order time discretization of the momentum equation (10). A correct solution of Eq. (10) for the pressure correction zeroes the grid divergence of $\boldsymbol{v}^{(n+1)}$ in Eq. (12). Performing of one time step requires solution of one Helmholtz equation for the temperature, three Helmholtz equations for the velocity components and one Poisson equation for the pressure.

The discretized Helmholtz and Laplace operators are inversed by the direct TPT method [32], which combines the eigenvalue decomposition of an operator with the Thomas algorithm. This direct method is shown to consume less computational time than standard Krylov-subspace or multigrid iteration techniques when large Reynolds (Grashof) number flows are being calculated on fine grids [32]. Furthermore, since the method used for the Laplacian inverse in Eq. (11) is direct, it yields the solution within the machine accuracy. Formulation of the present numerical schemes ensures that the consecutive action of the grid gradient and divergence result in the grid Laplacian. All these together guarantee that the correction step (12) brings the grid divergence values to machine zero (see [25] for a mathematical proof). In the computations below the maximal absolute grid divergence value always was below $10^{-14}$.

The above splitting method is $O(\delta t^2)$ for velocity, as is proved in [33]. However, it is argued in [34] that owing to Eqs. (11) and (12) the numerical error in the continuity equation $\nabla \cdot \boldsymbol{v} = 0$ is of the order of $\delta t$. In the current computations, however, the values of $\delta p$ and $grad(\delta p)$ are computed to within the machine precision, so that the error term proportional to $\delta t$ vanishes. Therefore, the resulting time discretization scheme is $O(\delta t^2)$ also for velocity divergence.

The second method used for time integration applies the same second-order discretization in time and pressure/velocity coupled Uzawa-like scheme proposed in [32]. This approach is based on LU decomposition of the Stokes operator

$$\begin{bmatrix} H_u & 0 & 0 & -\nabla_p^x \\ 0 & H_v & 0 & -\nabla_p^y \\ 0 & 0 & H_w & -\nabla_p^z \\ \nabla_u^x & \nabla_v^y & \nabla_w^z & 0 \end{bmatrix} = \begin{bmatrix} I & 0 & 0 & 0 \\ 0 & I & 0 & 0 \\ 0 & 0 & I & 0 \\ \nabla_u^x H_u^{-1} & \nabla_v^y H_v^{-1} & \nabla_w^z H_w^{-1} & I \end{bmatrix} \begin{bmatrix} H_u & 0 & 0 & -\nabla_p^x \\ 0 & H_v & 0 & -\nabla_p^y \\ 0 & 0 & H_w & -\nabla_p^z \\ 0 & 0 & 0 & C \end{bmatrix}, \quad (13)$$

$$C = \nabla_u^x H_u^{-1} \nabla_p^x + \nabla_v^y H_v^{-1} \nabla_p^y + \nabla_w^z H_w^{-1} \nabla_p^z. \quad (14)$$

Here $\nabla^x, \nabla^y$ and $\nabla^z$ are the first derivatives in the $x$-, $y$- and $z$- directions and $H = Gr^{-1/2}\Delta - \frac{3}{2\delta t}I$ are the Helmholtz operators, $\Delta$ is the Laplacian operator, $I$ is the identity operator and $\delta t$ is the time step. As above, the coefficient 3/2 follows from discretization of the time derivative. The lower indices show on which variable an operator acts, so that differences of the staggered grids, as well as possible different boundary conditions for different velocity components can be taken into account. The left hand side is a 4×4 operator matrix that assembles the 3D Stokes operator. The matrix $C$ is generalization of the Uzawa matrix. Unlike the Uzawa matrix, $C$ is not necessarily symmetric and not necessarily positive semi defined [32]. For this time integration technique the energy equation (9) is treated as above in Eq. (9), so that at each time step $T^{(n+1)}$ is computed first. The semi-implicit time step for the momentum and continuity equations is reduced to solution of the linear equations system

$$\begin{bmatrix} H_u & 0 & 0 & -\nabla_p^x \\ 0 & H_v & 0 & -\nabla_p^y \\ 0 & 0 & H_w & -\nabla_p^z \\ \nabla_u^x & \nabla_v^y & \nabla_w^z & 0 \end{bmatrix} \begin{bmatrix} u^{(n+1)} \\ v^{(n+1)} \\ w^{(n+1)} \\ p^{(n+1)} \end{bmatrix} = \begin{bmatrix} R_u \\ R_v \\ R_w \\ 0 \end{bmatrix} \quad (15)$$

where right hand sides $R$ contain the non-linear terms and all other terms that are treated explicitly. Using the LU decomposition (13), the solution is obtained in three steps:

1. Solve $\hat{u} = H_u^{-1} R_u$, $\hat{v} = H_v^{-1} R_v$ and $\hat{w} = H_w^{-1} R_v$ for $\hat{u}, \hat{v}$ and $\hat{w}$.
2. Solve $p^{(n+1)} = -C^{-1}(\nabla_u^x \hat{u} + \nabla_v^y \hat{v} + \nabla_w^z \hat{w})$ for $p^{(n+1)}$. \quad (16)
3. Solve $u^{(n+1)} = \hat{u} + H_u^{-1}\nabla_p^x p^{(n+1)}$, $v^{(n+1)} = \hat{v} + H_v^{-1}\nabla_p^y p^{(n+1)}$, and
$w^{(n+1)} = \hat{w} + H_w^{-1}\nabla_p^z p^{(n+1)}$.

Thus, carrying out of a single time step requires 6 inverses of the Helmholtz operator and one inverse of the pressure matrix $C$. It is worth noting that in this approach the pressure is defined only at the nodes lying inside the flow region (not boundary nodes), so that no pressure boundary conditions are needed. In the following the Helmholtz operators are inversed using the TPT method. Inverse of the matrix $C$ requires use of an iterative method. Our numerical experiments showed that the fastest convergence is obtained using ORTHOMIN(2) method [35]. It is worth mentioning that within our two approaches, the vectors $v^{(n+1/2)}$ obtained from Eq. (10) coincides with the vector $(\hat{u}, \hat{v}, \hat{w})$ calculated at the first stage of the algorithm (16).

## 4. Test calculations

At the preliminary stage of test calculations it was ensured that both time integration techniques described above arrive to the same steady state flow at subcritical Grashof numbers. These steady flows coincide with those obtained in [36] using a multigrid pressure-velocity coupling technique for the same finite volume discretization in space, that were also successfully compared with other independent results. In this section we focus on the grid convergence of the calculated slightly supercritical oscillatory flows, as well as establish that two independent time-integration techniques arrive to the same asymptotic oscillatory states. Visualization of three-dimensional velocity patterns is discussed in the next Section.

The grid convergence is established by comparison of the time histories of the total flow kinetic energy $E_{kin}$ and Nusselt number $Nu$ at the heated wall. These test calculations are illustrated in Fig. 1 and Table 1. The calculations were performed with the dimensionless time step 0.002 and remained unchanged when it was decreased to 0.001. Basing on this we assume the time step independence of the results and use $\delta t = 0.002$ for all the further calculations. Note that studies [14,15] reported time-step independence of the results starting from $\delta t = 1/16 = 0.0625$. Our numerical experiments show that this time step is too large for accurate computation of the critical parameters. Thus, for the CC – AA case, computations with $\delta t = 0.01$ yield $Gr_{cr} = 2.99 \cdot 10^6$. Reduction of the time step in ten times yields $Gr_{cr} = 3.30 \cdot 10^6$, which already seems to be time-step independent. It is noteworthy, that the oscillations frequency slightly above the critical point is 0.266 and 0.264 for the two above computations, respectively. The latter shows that the frequency scales with the $\sqrt{Gr}$, rather than a good agreement between the two results. Note also, that the time step in similar time-dependent calculations [19] was $5 \cdot 10^{-4}$, which is consistent with the current study.

Frame (a) of Fig. 1 compares oscillatory regimes computed for CC – AA case at $100^3$, $150^3$, and $200^3$ grids and shows that the results are almost visually indistinguishable when are obtained on the two finer grids, while oscillations at the $100^3$ grid slightly deviate at the minimal $E_{kin}$ and $Nu$ values. The corresponding rows of Table 1 show that $E_{kin}$ and $Nu$ average values converge up to the fourth significant digit, while for the oscillation amplitudes we can expect convergence within two first decimal places.

Case (b) of thermally insulated horizontal boundaries is significantly more challenging, because the boundary conditions allow for development of thin boundary layers along the vertical walls and the critical Grashof number becomes two orders of magnitude larger than that for the perfectly conducting horizontal boundaries [2,7,21]. Frames (b) of Fig. 1 show that results for $Gr=1.3 \cdot 10^8$ in the AA – CC case still remain slightly different for $200^3$ and $256^3$ grids, so that we cannot expect a complete convergence of the critical parameters there. Last four rows of Table 1 show that in spite of a noticeably larger Grashof number, the oscillation amplitudes are an order of magnitude smaller than in the previous case. This indicates additionally on the computational difficulties arising in the case of thermally insulated horizontal boundaries. The amplitudes convergence in the considered case is non-monotonic, and only the first decimal place is expected to be converged for the finest $256^3$ grid. At the same time the average values, as in the previous case, converge within the forth decimal place.

Figure 2 compares time-dependent calculations made by the pressure / velocity coupled and segregated solvers for two-frequency oscillatory regime obtained at quite large Grashof number $Gr=3 \cdot 10^8$ for AA – AA case (see below). This observation, showing that two independent solvers arrive to the same oscillatory regimes, allows us to rely on results obtained by the faster pressure / velocity segregated solver.

## 5. Visualization of three-dimensional flows

Three-dimensional temperature field can be easily represented by its isosurfaces. As an example, steady state temperature fields corresponding to subcritical flow states are shown in Fig. 3. One observes qualitative differences in the temperature patterns appearing due to different thermal boundary conditions. Owing to this observation, one would expect different most unstable disturbances and/or different instability development in each of the four cases considered.

Visualization of three-dimensional velocity fields is a considerably more complicated task. In the present study we implement the visualization method proposed in [24,25], making divergence-free projections of velocity on three sets of coordinate planes, (*x*,*y*), (*y*,*z*), and (*x*,*z*). Namely, we compute three projections $v_1$, $v_2$, $v_3$ of the velocity field $v$ on subspaces formed by divergence free velocity fields having only two non-zero components. Consider for example coordinate planes (*x*,*z*), a subspace formed by all vectors $a$, such that $a = [a_x, 0, a_z]$ and $\nabla \cdot a = \partial a_x/\partial x + \partial a_z/\partial z = 0$, and denote projection of the velocity field on this subspace as $v_1$. Similar projections on the subspaces defined in planes (*y*,*z*) and (*x*,*y*) are $v_2$ and $v_3$, respectively. Thus, we obtain three vector fields $v_1$, $v_2$, $v_3$, so that each of them has only two non-zero components, and each component is a three-dimensional scalar function. Namely,

$$v_1 = \begin{bmatrix} u_1(x,y,z) \\ 0 \\ w_1(x,y,z) \end{bmatrix}, \quad v_2 = \begin{bmatrix} 0 \\ v_2(x,y,z) \\ w_2(x,y,z) \end{bmatrix}, \quad v_3 = \begin{bmatrix} u_3(x,y,z) \\ v_3(x,y,z) \\ 0 \end{bmatrix}. \tag{17}$$

The two-dimensional divergence of each vector field vanishes:

$$div(v_1) = div_{(x,z)}(v_1) = \frac{\partial u_1}{\partial x} + \frac{\partial w_1}{\partial z} = 0 \tag{18}$$

$$div(v_2) = div_{(y,z)}(v_2) = \frac{\partial v_2}{\partial y} + \frac{\partial w_2}{\partial z} = 0 \tag{19}$$

$$div(v_3) = div_{(x,y)}(v_3) = \frac{\partial u_3}{\partial x} + \frac{\partial v_3}{\partial y} = 0. \tag{20}$$

This allows one to define a vector potential for each of the vectors, so that each vector potential has only one non-zero component,

$$v_1 = rot(\Psi_1); \quad \Psi_1 = (0, \Psi_y(x,y,z), 0) \tag{21}$$

$$v_2 = rot(\Psi_2); \quad \Psi_2 = (\Psi_x(x,y,z), 0, 0) \tag{22}$$

$$v_3 = rot(\Psi_3); \quad \Psi_3 = (0, 0, \Psi_z(x,y,z)) \tag{23}$$

Apparently, the three-dimensional function $\Psi_y(x,y,z)$ coincides with the stream function of the vector $v_1$ in each plane $y = const$. Therefore, $\Psi_y$ can be interpreted as an extended stream function, and the same can be said about $\Psi_x$ and $\Psi_z$. As a result, the fields $v_1, v_2$ and $v_3$ are tangent to the corresponding vector potential isosurfaces, and can be interpreted as divergence free projections of the velocity field on the coordinate planes. Arguments for uniqueness of these projections and different methods to compute them are given in [24,25].

Examples of this visualization showing three vector potentials together with the divergent-free velocity projections (depicted by arrows) on the corresponding coordinate planes are presented in Fig. 4. It is clearly seen that the divergence free velocity projection vectors are tangent to the isosurfaces of the vector potentials. It should be noticed also that projections on the (*x,z*) planes (left frames in Figs. 4a-4d) correspond to two-dimensional convective circulations altered by the three-dimensional effects. The comparison of 2D and 3D velocity patterns for the AA – AA case is reported in [24]. The three-dimensional effects are rather clearly seen from two other frames. The flow contains two pairs of diagonally symmetric rolls in the (*y,z*) planes (middle frames), and two other diagonally symmetric rolls in the (*x,y*) planes (right frames). Owing to motion along these rolls the main circulation depicted in the two right frames deviates from its two-dimensional counterpart. As is shown in [24], the deviation increases with the increase of the Grashof (or Rayleigh) number. Below we use this approach to visualize steady, time-averaged and time-dependent flows.

## 6. Transition to unsteadiness

A straight-forward way to find a critical Grashof number from a series of time-dependent computations is extraction of growth or decay rates from the time histories followed by the interpolation of growth rates to the zero value. This approach assumes that single-frequency oscillations amplitude monotonically grows or decays in time. In these cases, assuming a supercritical bifurcation, the critical Grashof number is obtained by a linear interpolation between two closest to zero negative and positive growth rates. If the bifurcation is found to be subcritical, the extrapolation to zero is done using two smallest decay rates. In some cases we observed two-frequency oscillatory regimes with amplitudes of either frequencies decaying, or one of them growing. To ensure correct results also for these cases we apply a non-linear least square fit that approximates a two-frequency signal, from which we extract growth/decay rates of each harmonics.

Also, as will be clear from the results below, it is essential to ensure that computations converge to a steady state, so that a slowly developing transient regime will not be mistakenly interpreted as a stationary state. Here, the computations process was terminated after the pointwise relative difference between all the variables at two consecutive time steps was below $10^{-9}$.

The results of interpolating of all the growth rates to zero and resulting critical Grashof numbers are reported in Table 2. The corresponding oscillation frequencies are given in Table 3. The results in Tables 2 and 3 are compared with the experiment [13], numerically converged 2D result of [22] obtained on $400^2$ grid, and several independent three-dimensional numerical results.

Cases with thermally insulated horizontal boundaries (AA – AA and AA – CC) are considerably more complicated and computationally challenging. Note the obvious scatter in the earlier results, obtained on coarser grids, reported in Tables 2 and 3. In all previous studies the number of grid or collocation nodes in one spatial direction did not exceed 100, which can be insufficient already for resolving thin thermal boundary layers adjacent to the cold and hot walls in a laterally heated two-dimensional square cavity. Also, in the present calculations we obtain qualitatively correct steady-oscillatory transitions only starting from $150^3$ grid. Calculations on $100^3$ grid sometimes yield wrong oscillations frequency since the most unstable perturbation is not yet resolved numerically. We assume that insufficient spatial resolution is the main reason for disagreement in the previous results.

Flow time histories are monitored by storing the values of dimensionless total kinetic energy $E_{kin}$ and Nusselt number $Nu$ at the cold wall, which are defined as

$$E_{kin} = \frac{1}{2}\int_V (u^2 + v^2 + w^2)dV \tag{24}$$

$$Nu = \int_0^1 \int_0^1 \left[\frac{\partial T}{\partial x}\right]_{x=0} dydz \tag{25}$$

Since all the oscillatory flows computed at small supercriticalities are periodic, the corresponding average flows can be easily calculated by integration over a single period. At the same time, since the oscillation frequency $\omega$ is already known via Fourier analysis of the time dependencies of the total kinetic energy and the Nusselt number, additional time integration over a period can be performed together with evaluation of the Fourier integral

$$\{A_u, A_v, A_w, A_T\}_{ijk} = \frac{\omega}{\pi}\int_0^{2\pi/\omega}\{u, v, w, T\}_{ijk} e^{-i\omega t} dt \tag{26}$$

where $A$ with a subscript stays for the amplitude of the corresponding function oscillations with the frequency $\omega$, and the integral is performed at each grid point. The three-dimensional amplitude functions describe the spatial pattern of the oscillations amplitude for each scalar function. When the bifurcation is supercritical and supercriticality is small they also represent patterns of the most linearly unstable disturbance. In the following we use the absolute value of $A_T$ to illustrate calculated oscillatory regimes.

The slightly supercritical oscillatory flows are visualized by snapshots of the vector potentials described in [24,25] and illustrated in Fig. 4. Animations of the oscillatory regimes described below are supplied as separate files, and are reported as 3D snapshots. We also tried to compare time-averaged flow patterns with those of the steady states. Steady states at the supercritical Grashof numbers were calculated by the Newton method using the approach of [37] with the large-time step iterations performed as proposed in [32].

*6.1 CC-CC and CC-AA cases*

The amplitude growth and further non-linear evolution of the flow oscillations for the CC-CC and CC-AA cases is illustrated in Fig. 5. The steady-oscillatory transition seems to be similar: in both cases the exponential amplitude growth is followed by a transition to another oscillatory regime (see below).

As follows from Table 2, in both CC-CC and CC-AA cases the critical Grashof number and the critical frequency are almost independent on the spanwise boundary conditions. The obtained value of $Gr_{cr} \approx 3.3 \cdot 10^6$ is well compared with the two-dimensional one, as well as with the results of [14,15], however is noticeably lower than the experimentally observed value $4.5 \cdot 10^6$ of [13]. The discrepancy can be caused, for example, by imperfections of experimental boundary conditions, or by sensitivity of thermocouples that could measure temperature oscillations starting from rather large amplitude. The oscillation frequencies are well compared with the two-dimensional prediction, and are close to the experimentally measured one.

Amplitude of the temperature oscillations at $Gr=3.4 \cdot 10^6$ is shown in Fig. 6a,b for the CC-CC and CC-AA cases, respectively. The amplitude isosurfaces are colored by values of the time averaged temperature at the same points. Both amplitude patterns are similar. We observe that the maximal amplitudes are located in the lower left and upper right corners of the midplane $y=0.5$. Note that in the lower left corner the hot fluid is located below the cold one, while in the upper right corner the cold fluid is located above the hot one, as is indicated by the colors. This observation was made previously in [14,15], where it was argued that the instability is caused by the Rayleigh-Bénard mechanism. Figure 6c shows isotherms and distribution of the amplitude of the most unstable perturbation obtained by the linear stability analysis [7,22] for the two-dimensional CC case. Comparison with the 3D result, which must be done in the midplane $y = 0.5$, shows that in the 2D configuration we observe a similar pattern of the amplitude with the maxima located in the lower left and upper right corners. Taking into account close critical Grashof numbers and critical frequencies (Tables 2 and 3), we conclude that in these cases the 2D model correctly describes the instability onset, which takes place due to the Rayleigh-Bénard mechanism.

Comparing the subcritical steady states at $Gr=3.2 \cdot 10^6$ (Fig. 4a,b) we observe that in the CC – CC case the flow is more intensive. The latter follows from comparison of maximal values of the vector potentials that can be interpreted as extended stream functions.

Comparison of the maximal velocity values yields for $v_x$, $v_y$, and $v_z$, respectively, 0.251, 0.107, and 0.371 in the CC – CC case; and 0.228, 0.0563, and 0.342 CC – AA case. Apparently, this difference in flow intensities is caused by different temperature distribution near the spanwise boundaries (Fig. 3), and, surprisingly, does not affect the instability threshold, which is entirely two-dimensional, so that the most intensive disturbances are located near the $y = 0.5$ midplane. However, it affects further instability development. Thus, examination of slightly supercritical flows at $Gr$=3.4·10$^6$ shows that the oscillatory flow in the CC – CC case preserves the reflection symmetry, while 2D rotational symmetry and centro-symmetry are broken. In the CC – AA case all the three symmetries are broken at $Gr$=3.4·10$^6$. Note that the 2D instability sets in with the break of rotation symmetry, which follows from rotation-antisymmetric pattern of its most unstable perturbation, whose absolute value is shown in Fig. 6c [7]. Thus, break of rotation symmetry is another similarity between 2D and 3D models with perfectly conducting horizontal boundaries.

If the steady – oscillatory transition take place via a supercritical Hopf bifurcation and results into a stable limit cycle, the time-averaged flow remains close to the unstable steady state one (see e.g., [38]). Comparison of the total kinetic energy and the Nusselt number of unstable steady state and oscillatory flows at $Gr$=3.4·10$^6$ results in the following. In the CC – CC case $E_{kin}$ =6.666·10$^{-3}$ and 6.669·10$^{-3}$, $Nu$ =7.591 and 7.600, for the steady and oscillatory states, respectively. In the CC – AA case the same comparison yields $E_{kin}$ =4.796·10$^{-3}$ and 4.890·10$^{-3}$, $Nu$ =7.481 and 7.514. Thus, these two integral characteristics remain close in both cases, which allows us to assume that we observe supercritical Hopf bifurcations resulting in stable limit cycles. A larger difference in the CC – AA case can be due to steeper increase of the growth rate with the growth of the Grashof number.

The time averaged flow patterns corresponding to oscillations shown in Fig. 5 almost do not differ from that of the subcritical steady state (Fig. 4), at least up to $Gr$=3.4·10$^6$. In both cases the flow oscillations are noticeable for the potentials $\Psi_y$ that describes the main convective circulation (Fig. 7) and $\Psi_x$ (Fig.8). For comparison, patterns of $\Psi_y$ of unstable steady states at $Gr$=3.4·10$^6$ are also included in Fig. 7. We observe that circulating flow around the cavity persists only near the sidewalls, while in the central part motion in the (*x,z*) planes becomes quite complicated. In the two cases considered it differs in the central part of the cube, but this difference is mostly a result of different steady state flow patterns. Examining oscillations of the potential $\Psi_x$ describing motion in (*y,z*) planes, we observe that it remains almost unchanged in the CC – CC case, while oscillates noticeably in the CC – AA case.

Trying to follow development of supercritical oscillatory regimes at larger Grashof numbers, we observe that in the CC – CC case the oscillations take place with a single frequency, closed to one reported in Table 3, and its harmonics. In the CC – AA case the oscillatory regime becomes unstable already at $Gr=3.4·10^6$ and transforms into a multi-frequent regime with at least five dominating frequencies.

*6.2 AA – CC case*

In the AA-CC case, starting from $150^3$ grid, we observe oscillations with a low dimensionless frequency, $f_{cr}≈0.01$ that appear at $Gr_{cr}≈1.2·10^8$. The calculations in this, as well as in the next AA – AA case, are much more computationally demanding, not only because of the noticeably larger Grashof number, but also because of low-frequency oscillations that require very long computational times to arrive to an asymptotic oscillatory regime. Since all previous studies for thermally insulated horizontal boundaries were done for the AA – AA case only, no independent data is available for comparison. It is stressed again that calculations on a coarser $100^3$ grid resulted in oscillations with a quite different frequency and also showed transitional time histories different from those discussed below.

Time histories of the total kinetic energy at slightly sub- and super-critical Grashof numbers are shown in Fig. 9a. Being in the subcritical regime, at $Gr=1.1·10^8$ and $1.2·10^8$, we observe several decaying irregular oscillations that after some time result in a single-frequency oscillations with decaying amplitude. After the Grashof number is increased beyond the critical one, the initial irregular oscillations transform into a many-frequent periodic regime containing the main frequency and its three to five harmonics, as is shown in Fig. 9b. A regular transient part of this transition lasts only 3 – 4 periods, so that we were unable to extract growth rates from these signals. The critical values reported in Table 2 are obtained by extrapolation of decay rates to zero, and the frequencies reported in Table 3 are taken from the decaying time histories. The oscillations pertain for a long time, about 20 – 30 oscillation periods, without any evidence of instability of this limit cycle. The amplitude of oscillations remains relatively small, so that the Nusselt number and the total kinetic energy of the time-averaged and unstable steady flow at $Gr=1.3·10^8$ differ only in the third decimal place. Reduction of the Grashof number from $1.3·10^8$ to $1.25·10^8$ resulted in a steady state flow. Thus, we can assume that this is a supercritical bifurcation into a many-frequent limit cycle.

The transition observed breaks both rotation and 2D reflection symmetries, but sustains the centro-symmetry of the flow. Thus, examination of the temperature distribution in a certain snapshot shows that the maximal absolute deviation between $\theta(x,y,z)$ and either of $\theta(x, 1-y, z)$ or $-\theta(1-x, y, 1-z)$ is exactly 0.0632. The maximal absolute deviation between $\theta(x,y,z)$ and $\theta(1-x, 1-y, 1-z)$ is $1.9 \cdot 10^{-7}$. Here we observe the exceptional case described in Section 2.

Pattern of the temperature amplitude in a slightly oscillatory regime at $Gr=1.3 \cdot 10^8$ is shown in Fig. 10. The amplitude isosurfaces (Fig. 10a) form many small-scale structures that conceal most of its important properties. To get more insight, the amplitude isolines are plotted in characteristic cross-sections by the coordinate planes (Fig. 10b,c,d). We observe that there are no intensive temperature oscillations in the central part of the box. The maximal amplitude values (red color) are located near all the six box borders, namely, (i) near the spanwise boundaries $y=0$ and 1 in the corners $x=z=1$ and $x=z=0$, respectively (Fig. 10b); (ii) near the isothermal boundaries $x=0$ and 1 in the corners $y=z=0$ and $y=z=1$, respectively (Fig. 10c); and (iii) in the upper and lower parts of the cavity inside the boundary layers adjacent to the vertical boundaries (Fig. 10d). Clearly, such amplitude pattern cannot be directly compared to a 2D result for the AA square cavity (see below), since similarity, if any, is expected to be observed in the midplane y=0.5. We conclude that the oscillations observed have fully three-dimensional origin.

Comparison of the potential isosurfaces of unstable steady state and time-averaged flow (Fig. 11) at $Gr=1.3 \cdot 10^8$ does not reveal any noticeable qualitative changes in the potentials $\Psi_y$ and $\Psi_x$. The potential $\Psi_z$ changes noticeably in the central part of the box. This change is illustrated additionally by arrow plots in the *z=const* planes where we observe break of reflection symmetry of vortical motion in the *(x,y)* planes, which is caused by oscillations of the z-component of vorticity. We observe also the fastest convective motion along the main circulation (left frames) shifted towards the spanwise boundaries. Comparing maximal values of the vortex potentials, we notice also that in the time-averaged flow the main circulation slightly weakens, compared to the steady state one, while the motion in the two remaining coordinate planes slightly intensifies. The oscillations snapshots (not shown in the figures) reveal that the most noticeable oscillations are located in the central part of the cube, while the motion remains fastest near the spanwise boundaries.

*6.3. AA – AA case*

This case appears to be the most difficult for computational simulation. Previous computations performed on coarser grids reported transition to unsteadiness at the Grashof number between $4·10^7$ and $5·10^7$ (Table 2). At the same time our previous computations on $100^3$ grid [24] arrived to a steady flow at $Ra=GrPr=10^8$, which was verified here by calculations performed on the three finer grids, and agrees with the result of [20]. To resolve this apparent contradiction we needed to perform a series of very careful CPU-time consuming computations varying the Grashof number from $4·10^7$ to $3·10^8$. The steady state flow patterns of this case were already reported in [24] and are not shown here.

The first steady – oscillatory transition was found to take place at $Gr_{cr}^{(1)} \approx 4.6·10^7$ with breaking of all the three symmetries. The characteristic time dependence of the total kinetic energy and the Nusselt number is shown in Fig. 12. The calculations start from the steady state at $Gr=4.55·10^7$. An abrupt increase of the Grashof number to $4.6·10^7$ results in irregular large-amplitude oscillations that are attributed to non-modal disturbances growth mechanism. These oscillations decay in a relatively short time, after which the flow remains almost unchangeable during ≈500 time units, so that the pointwise relative difference between the two time steps becomes of the order of $10^{-7}$. The latter can be mistakenly interpreted as a steady state. As mentioned above, to ensure that computations converge to a steady state (e.g., at $Gr=4.5·10^7$), and to distinguish it from a slowly developing transient regime, the computational process was terminated only after the pointwise relative difference between two consecutive time steps was below $10^{-9}$. After a long transient period, the amplitude of oscillations starts to grow, and finally the computational process arrives to an oscillatory flow regime. The spectrum of the oscillations shown in the insert of Fig. 12 contains the main frequency $f \approx 0.01$, its harmonics and a sub-harmonic. Clearly, much longer time integration is needed to make a definite conclusion about the asymptotic oscillatory state.

A long transient time needed for the instability development can be connected to the symmetry breaking. Since initial state is symmetric, no non-symmetric perturbations are introduced. The numerical growth of these perturbations starts from the round-off errors and may consume a long time until arriving to noticeable amplitude that still remains relatively small compared to other cases considered. Note that similar calculations in [16] resulted in the break of only reflection symmetry with preserved rotational one. Possibly, in these calculations, the break of rotation symmetry was not resolved, which resulted in a critical Grashof number larger than the one computed here.

The calculations performed for $Gr$=4.55·10$^7$ resulted in a steady state when were started from the steady flow at $Gr$=4.5·10$^7$. When the oscillatory state at $Gr$=4.6·10$^7$ is taken as an initial condition, the computations at $Gr$=4.55·10$^7$ arrived to an oscillatory flow. This indicates on the sub-critical steady-oscillatory transition.

The second transition occurs when the Grashof number crosses the second critical value $Gr_{cr}^{(2)} \approx 7.1\cdot10^7$. With increase of the Grashof number already to $Gr$=7.15·10$^7$ the oscillatory flow turns into stable and steady one with all the symmetries reinstated. Figure 13 illustrates destabilization of the flow with decrease of the Grashof number from 7.2·10$^7$ to 7.0·10$^7$. We observe that being steady and stable at $Gr$=7.15·10$^7$, the flow become unstable at $Gr$=7.1·10$^7$. Computations at $Gr$=7·10$^7$ and $Gr$=7.1·10$^7$ (Fig. 13) exhibit first quasi-steady and then quasi-regular oscillatory time-dependencies. At longer times the oscillations become irregular. Significantly longer time integration is needed to make a conclusion about stochastic properties of their asymptotic states, which is beyond the scope of the present study.

The restored stable steady states remain stable up to $Gr_{cr}^{(3)} \approx 2.8\cdot10^8$. As is shown in Figs. 14 and 15, the oscillations at $Gr$=2.7·10$^8$ slowly decay, while at $Gr$=2.8·10$^8$ the time dependencies become more complicated. At the transitional stage we observe decaying low-frequency oscillations that are modulated by larger frequency ones. The spectrum of these oscillations, plotted in the insert of Fig. 14, shows that the two frequencies differ in more than an order of magnitude. Finally, only large frequency oscillations with very small amplitude remain. Gradually increasing the Grashof number, we observe that the oscillation amplitude slowly grows, as is expected (Fig. 15). The lower frequency, that modulates the larger-frequency time dependence (Fig. 6), varies from 0.043 at $Gr$=2.8·10$^8$ to 0.090 at $Gr$=3·10$^8$. This transition preserves the three symmetries that start to break already at $Gr\approx$2.9·10$^8$. This time behavior was observed for 150$^3$, 200$^3$ and 256$^3$ grids.

As follows from Table 2, calculations on 100$^3$ grid for the AA – AA case fail to predict either correct critical Grashof number, or correct frequency of oscillations, or both. A similar conclusion can be made when current fine grid results are compared with the previous studies performed on coarser grids. Thus, critical Grashof numbers and oscillation frequencies reported in [16,17,19] coincide with the present result for the first bifurcation, as well as between themselves, only in the first decimal place. Results of [18,20] differ drastically from the above cited ones, and possibly relate to the third transition reported here. Their critical Grashof numbers are of the same order of magnitude as our $Gr_{cr}^{(3)}$. The oscillations frequency reported in [20] is 0.532, which is close to the present result. The dimensionless frequency obtained in [18] and scaled by $H^2/\alpha$, is 0.43. If it is rescaled by $Pr/\sqrt{Gr}$ to be compared with the present results, it becomes surprisingly small.

Amplitude of the temperature oscillations after the first transition at $Gr_{cr}^{(1)} \approx 4.6 \cdot 10^7$ is shown in Fig. 16a-c, and is compared with the most unstable temperature perturbation in the 2D cavity with perfectly insulated horizontal boundaries (AA cavity). We observe sharp maxima of the amplitude in the lower right and upper left corners with much weaker maxima in the opposite corners (Fig. 16a). Some more insight can be gained also from cross-sectional plots in Fig. 16b. Since the large amplitude values are located near the midplane $y=0.5$ and the global maxima lay in the midplane, we can compare the pattern with the most unstable perturbation of the corresponding 2D instability in the square cavity. Thus, comparing Fig. 16c and 16d we observe that the maxima of both cases are located in the same corners, however patterns of amplitudes noticeably differ. Taking into account that steady-oscillatory transition in the 2D case takes place at significantly larger Grashof number, $Gr \approx 2.6 \cdot 10^8$, we can state only partial similarity of the 2D and 3D instabilities. Stability of the 2D flow in a square cavity with respect to spanwise-periodic three-dimensional perturbations was studied in [23]. The transition from steady 2D to steady 3D flow was found at $Gr=2.18 \cdot 10^7$, which is completely different result compared to the present one. Nevertheless, the temperature perturbation shown in Figure 1 of [23] exhibit some similarity with our shown in Fig. 16a.

Due to the small oscillations amplitude the time-averaged and steady-state flow patterns look undistinguishable and are not shown. As previously, basing on this small difference, we assume that the bifurcation is super critical. The oscillations consists of weak pulsations of the main convective vortex (potential $\Psi_y$) and stronger oscillations in two other coordinate planes (potentials $\Psi_z$ and $\Psi_x$), as is seen from the animation at $Gr=5 \cdot 10^7$ attached to Fig. 16.

Fourier analysis of the time dependencies slightly below the second oscillatory –steady transition (Fig. 13) shows three frequency peaks for dimensionless frequencies 0.00610, 0.0122, and 0.0183, with the largest peak corresponding to the second value. It is not clear whether the two larger frequencies are harmonics of the lower one, or the lower one results from a period doubling of the second frequency. The time averaged flow calculated at $Gr=7 \cdot 10^7$ is very close to the corresponding unstable steady state (Fig. 17). As above, this indicates on a smooth transition from steady to oscillatory regime with the decrease of Grashof number. Amplitudes of the three frequencies are plotted in Fig. 18. All of them have similar patterns that are qualitatively different from one observed at the first transition (Fig.

16). Recalling the linear stability analysis, the similarity indicates again that oscillations at all the three frequencies result from the same eigenmode of the linearized problem.

When the Grashof number is increased slightly above $2.8 \cdot 10^8$ we observe appearance of two-frequency oscillations (Fig. 14). As mentioned above, integration over a longer time leads to a decay of the lower frequency oscillations, so that only low-amplitude and large-frequency oscillations sustain, as is shown in Fig. 15. The characteristic frequency of these oscillations is 0.525 (Table 3). This regime remained stable and preserved all the symmetries for long integration times when computations were performed on $200^3$ and $256^3$ grids. When the calculations started from the developed oscillatory regime at $Gr=2.8 \cdot 10^8$ and the Grashof number was reduced to $Gr=2.7 \cdot 10^8$, the oscillatory regime sustained, from which we concluded that the observed instability is sub-critical.

Due to very small oscillations amplitude the time-average and unstable steady flow regimes are very close and look undistinguishable on a graph (not shown here). The amplitude of the temperature oscillations of this regime is shown in Fig. 19a. Its pattern can be seen as a superposition of the patterns resulting from the two previous transitions (Figs. 16 and 18). We observe that the largest amplitudes are located inside rotation-symmetric small-scale structures in the lower right and the upper left corners, which is similar to the pattern of the first transition (Fig. 16). Two other rotation-symmetric structures with smaller amplitudes start from the opposite corners and propagate deeper into the flow similarly to the amplitude patterns of the second transition (Fig. 18).

Computations on $150^3$ grid at $Gr=2.9 \cdot 10^8$, starting from a snapshot at $Gr=2.8 \cdot 10^8$, arrived to the above described two-frequency regime, which became unstable, and after a longer time integration resulted in another oscillatory regime with all the symmetries broken. The corresponding history of the total kinetic energy and the Nusselt number is shown in Fig. 20. The new (second) oscillatory regime is characterized by three frequencies 0.024, 0.524 and 0.549 clearly seen in the Fourier spectrum. Using an interpolated snapshot of this regime as an initial state for calculation on finer grids, shows that it pertains almost unchanged, as is shown in Fig. 2b for $Gr=3 \cdot 10^8$. When Grashof number is increased to $3 \cdot 10^8$, the lowest frequency remains the same, while two larger frequencies change to 0.528 and 0.553.

We tried to reproduce the instability of the single frequency oscillatory regime, as is shown in Fig. 20, calculating on the two finer grids, and succeeded too observe it on $200^3$ grid at $Gr=3 \cdot 10^8$. On the $256^3$ grid the large frequency regime did not exhibit any instability for a very long integration time. Thus, it remains unclear whether the instability sets in at even a more longer time, or at a larger Grashof number, or two oscillatory regimes become separated at finer grids. Further, we performed calculations on $150^3$ and $200^3$ grid gradually reducing the Grashof number by decrements of $10^7$. Oscillations corresponding to the second oscillatory regime sustains up to $Gr=2.3\cdot10^8$, where the flow oscillates with the main frequency 0.03 and its harmonics.

To illustrate these oscillatory regimes by animation or snapshots is rather difficult since either oscillations amplitude is too small, or too many frames are needed to be stored or plotted. Some additional information can be derived from the spatial distribution of the frequencies amplitude and time-averaged flow pattern. Thus, Figure 20 compares spatial distribution of the oscillation amplitude of the primary large-frequency oscillations (Fig. 20a) with those of the three frequencies found in spectrum of the second oscillatory regime (Figs. 23b-23d). A rather surprising observation is that in the amplitude structures that appeared together in the single frequency regime become separated and each of them oscillates with its own frequency. This separation can be connected with the observed symmetry breaking.

To compare the 2D and 3D cases also for this transition, the amplitude of 3D temperature oscillations is plotted in several $y=const$ cross-sections together with the time-averaged isotherms. It is seen that the 3D amplitude is noticeably larger near the midplane $y=0.5$, which would indicate on some analogy with the 2D case. However, the shape of the 3D amplitude pattern is not similar to its 2D counterpart, and is located "deeper" in the corners (cf. Figs. 16b and 21). Also, presence of large (but not largest) amplitude oscillations propagating from the lower left and upper right corners in the bulk of the flow indicates on different, compared to the 2D case, origin of the instability.

## 7. Conclusions and discussion

Steady – oscillatory transition and slightly supercritical oscillatory states of buoyancy convection in a laterally heated cube were studied by the straight-forward time integration of Bousinesq equations. Pressure / velocity segregated and coupled time integration schemes were applied. Four different sets of thermal boundary conditions were taken into consideration. Either perfectly thermally conducting (CC) or perfectly thermally insulated (AA) horizontal and spanwise boundaries were considered. A special attention was paid on the grid convergence of the results. Three-dimensional velocity fields are visualized using a novel approach of quasi-two-dimensional divergence free projections [24,25]. The obtained values of the critical Grashof numbers and critical oscillation frequencies comprise a set of benchmark data needed for validation of future numerical methods that will allow one to solve fully three-dimensional (Tri-Global) stability problems. The results for CC – CC and AA – CC cases are new. The second and third transitions in the AA – AA case are quite unexpected, and were never reported before. The results for CC – AA case and for the primary bifurcation in the AA – AA case noticeably improve accuracy of the previous studies.

When cube horizontal boundaries are perfectly thermally conducting (CC – CC and CC – AA cases), the steady-oscillatory transition takes place at $Gr \approx 3.3 \cdot 10^6$, which, together with the calculated oscillation frequencies, agree well with the previous findings of [14,15], and is not far from the experimentally measured values of [13]. The critical Grashof numbers and the oscillations frequency are also close to those obtained for convection in a laterally heated two-dimensional square cavity [7,22]. We showed also that amplitude of the most unstable two-dimensional perturbation, resulting from the linear stability analysis, is similar to the three-dimensional pattern of the oscillations amplitude. All these allows us to argue that in the case of perfectly conducting horizontal walls the two- and three-dimensional instabilities set in owing to the same physical reasons and support argument made in [14,15], that this instability is driven by local Rayleigh-Bénard mechanisms. In both cases the steady – oscillatory transitions are super-critical. At the same time, in spite of the similar instability mechanism, the two bifurcations differ with respect to the symmetry breaking: in the CC – CC case the reflection symmetry is preserved, while in the CC – AA case all the symmetries are broken. Consequently, further flow changes, even at small supercriticalities, differ qualitatively.

When the horizontal boundaries are perfectly thermally insulated (AA – CC and AA – AA cases) the primary bifurcation takes place at Grashof numbers that are more than an order of magnitude larger than those obtained for the perfectly insulated horizontal boundaries. Also, both oscillations amplitude and frequency become about an order of magnitude smaller, which cause additional numerical difficulties. It was observed that the primary steady – oscillatory transition is qualitatively different for perfectly thermally conducting (AA – CC) and perfectly insulated (AA – AA) spanwise walls.

In the AA – CC case the critical Grashof number is found to be beyond $1.2 \cdot 10^8$, and oscillations appear with a relatively low dimensionless frequency $\approx 0.01$. The transition from steady to oscillatory regime is super-critical. No independent numerical or experimental data is available here for comparison. The instability observed does not exhibit any similarities with the corresponding 2D AA case.

In the AA – AA case three consecutive steady – oscillatory transitions were observed and two of them are reported here for the first time. The first one takes place at $Gr \approx 4.6 \cdot 10^7$ with the break of all the symmetries and via a sub-critical bifurcation. The critical Grashof number and oscillations frequency are close to previously reported values [17,18,19] and are converged to within the second decimal digit. At $Gr≈7.2·10^7$ the stability of steady states restores together with all the symmetries. We presented some arguments showing that this transition is super-critical with respect to decreasing Grashof number. Finally, at $Gr≈2.8·10^8$ the steady flow becomes unstable sustaining the symmetries. There is also some evidence that the resulting single frequency oscillatory flow becomes unstable again already at $Gr≈2.9·10^8$ and transforms into oscillations with three characteristic frequencies and a break of the symmetries. This transition indicates on possible sub-criticality, so that single and triple frequency regimes are observed at the same Grashof numbers.

Returning to the comparison of 3D and 2D results on the steady – oscillatory transition, we conclude that only CC – CC and CC – AA cases exhibit complete similarity with its two-dimensional CC counterpart. Only partial similarity is observed for the first bifurcation in the AA – AA case, while the AA – CC case no similarities were observed.

**Acknowledgement**

This work was supported by the LinkSCEEM-2 project, funded by the European Commission under the 7th Framework Program through Capacities Research Infrastructure, INFRA-2010-1.2.3 Virtual Research Communities, Combination of Collaborative Project and Coordination and Support Actions (CP-CSA) under grant agreement no RI-261600. The author acknowledges PRACE for awarding him access to resource CURIE based in France at Très Grand Centre de Calcul.

**Figure Captions**

Figure 1. Flow oscillations at $Gr=3.4 \cdot 10^6$, CC – CC case (a), and $Gr=1.3 \cdot 10^8$, AA – CC case (right), at different finite volume grids.

Figure 2. Comparison of time-dependent computations by pressure / velocity coupled and segregated solvers for two-frequency oscillatory regime calculated at $Gr=3 \cdot 10^8$, AA – AA case, $200^3$ grid.

Figure 3. Temperature isosurfaces corresponding to slightly subcritical steady states.

Figure 4. Visualization of 3D velocity fields corresponding to slightly subcritical steady states. Divergence free projections of velocity fields on the coordinate planes are shown by vectors. Isosurfaces of the velocity potentials, to which the vector fields are tangent, are shown by colors. In the frames (a), (b), and (c) the potential isosurfaces are plotted for the levels of 0.3 multiplied by the corresponding maximal absolute value. In the frame (d) the factors are 0.5, 0.3, and 0.4 for $\Psi_y$, $\Psi_x$, and $\Psi_z$, respectively..

Figure 5. Time evolution of the total kinetic energy and the Nusselt number from $Gr=3.4 \cdot 10^6$ to $Gr=3.5 \cdot 10^6$. (a) and (b) CC – CC case; (c) and (d) CC – AA case. Calculation on the $100^3$ grid.

Figure 6. Amplitude of temperature oscillations at $Gr=3.4 \cdot 10^6$ in CC – CC (a) and CC – AA (b) cases. Colors of the isosurfaces correspond to the average temperature values. (c) Isotherms (black lines) and amplitude of the most unstable temperature perturbation (colors) in the 2D case with perfectly conducting horizontal boundaries. Calculation on the $100^3$ grid. Maximal values of the amplitude and the levels plotted are, respectively, 0.0769, 0.01, 0.02, 0.03 in the frame (a) and 0.114, 0.015, 0.025, 0.035 in the frame (b).

Figure 7. Snapshots of isosurfaces of vector potential $\Psi_y$ corresponding to the main convective circulation compared to the patterns of unstable steady state flows. $Gr=3.4 \times 10^6$. The plotted levels are 0.00335 and 0.0223 for the CC – CC case and 0.0015 and 0.0195 for the CC – AA case. Calculation on the $100^3$ grid. Animation1. Animation2.

Figure 8. Snapshots of isosurfaces of vector potential $\Psi_x$ compared to the patterns of unstable steady state flows. $Gr=3.4 \times 10^6$. CC – CC case (left frames), and CC – AA case (right frames). The plotted levels are $\pm 0.0073$ for the CC – CC case and $\pm 0.004$ the CC – AA case. Calculation on the $100^3$ grid. Animation1. Animation2.

Figure 9. Time evolution of the total kinetic energy at $Gr=1.1 \cdot 10^8$ to $Gr=1.4 \cdot 10^8$ for the AA – CC case (a) and frequency spectra at two supercritical values of the Grashof number (b). Calculation on the $150^3$ grid.

Figure 10. (a) Isosurfaces of amplitude of temperature oscillations at $Gr=1.3 \cdot 10^8$ in AA – CC case. The maximal value and levels plotted are 0.0302, 0.004, 0.007, 0.01. (b), (c), and (d)

Amplitude isolines in characteristic $y$-, $x$-, and $z$- cross-sections respectively. Calculation on the $200^3$ grid. Animation 3.

Figure 11. Comparison of vector potentials of the time-averaged oscillatory flow states at $Gr=1.3 \cdot 10^8$ in AA – CC (upper frames) with those of the unstable steady state at the same Grashof number (lower frames). Isosurface are shown at $\Psi_y = 0.006, 0.01$ and $0.02$, $\Psi_x = \pm 0.004$, and $\Psi_z = \pm 0.004$. Calculation on the $200^3$ grid.

Figure 12. Time evolution of the total kinetic energy and Nusselt number at $Gr=4.6 \cdot 10^7$ starting from steady state at $Gr=4.55 \cdot 10^7$ in the AA – AA case. The insert shows the frequency spectra of both time dependencies calculation for $2000 < t < 3000$. Calculation on the $200^3$ grid.

Figure 13. Time evolution of the total kinetic energy from $Gr=7.2 \cdot 10^7$ up to $Gr=7.0 \cdot 10^7$ starting from steady state at $Gr=7.25 \cdot 10^7$ in the AA – AA case. Calculation on the $200^3$ grid.

Figure 14. Time evolution of the total kinetic energy from $Gr=2.6 \cdot 10^8$ to $Gr=2.7 \cdot 10^8$ (red line), then to $Gr=2.8 \cdot 10^8$ (green line), and then to $Gr=2.9 \cdot 10^8$ (blue line). AA – AA case. Calculation on the $200^3$ grid.

Figure 15. Large frequency / small amplitude oscillations of the Nusselt number computed for gradually increased Grashof number in AA – AA case. Calculation on the $200^3$ grid.

Figure 16. (a) Isosurfaces of amplitude of temperature oscillations at $Gr=4.7 \cdot 10^7$ in AA – AA case. (b) Amplitude of the temperature oscillations in several $y=const$ planes of the frame (a). (c) Isotherms (black lines) and amplitude of the temperature oscillations (colors) in $y=0.35$ plane of the frame (a). (d) Isotherms (black lines) and amplitude of the temperature oscillations (colors) of the most unstable perturbation in the 2D AA case. Calculation on the $150^3$ grid Animation 4.

Figure 17. Comparison of vector potentials of the time-averaged oscillatory flow states at $Gr=7 \cdot 10^7$ in AA – AA (upper frames) with those of the unstable steady state at the same Grashof number (lower frames). Isosurface are shown at $\Psi_y = 0.0045, 0.006$ and $0.0075$, $\Psi_x = \pm 0.0006$, and $\Psi_z = \pm 0.001$. Calculation on the $200^3$ grid.

Figure 18. Amplitudes of three-frequent temperature oscillations at $Gr=7 \cdot 10^7$, AA – AA case. The values of dimensionless frequencies are $0.00610, 0.0122$, and $0.0183$ for the frames (a), (b), and (c), respectively. Calculation on the $200^3$ grid.

Figure 19. Time evolution of the total kinetic energy (a) and Nusselt number (b) at $Gr=2.9 \cdot 10^8$. AA – AA case. Calculation on the $150^3$ grid.

Figure 20. Amplitudes of temperature oscillations at $Gr=2.9 \cdot 10^8$, AA – AA case. (a) amplitude of the single frequency regime shown in Fig. 15. (b) – (d) amplitudes of the three main frequencies of the second oscillatory regime shown in Figs. 2b and 17. Calculation on the $150^3$ grid.

Figure 21. Time-averaged isotherms (colors) and amplitude of the single frequency oscillatory regime at $Gr=2.9 \cdot 10^8$, AA – AA case. Calculation on the $150^3$ grid.

Table 1. Grid convergence of the average values and amplitudes of the total kinetic energy and the Nusselt Number for two characteristic cases illustrated in Fig. 1.

| Grid | CC – CC case, $Gr=3.4 \cdot 10^6$ | | | | AA – CC case, $Gr=1.3 \cdot 10^8$ | | | |
|---|---|---|---|---|---|---|---|---|
| | $\bar{E}_{kin}$ | $\overline{Nu}$ | $A(E_{kin})$ | $A(Nu)$ | $\bar{E}_{kin}$ | $\overline{Nu}$ | $A(E_{kin})$ | $A(Nu)$ |
| $100^3$ | $6.6949 \cdot 10^{-3}$ | 7.59988 | $6.306 \cdot 10^{-5}$ | 0.1935 | $2.12696 \cdot 10^{-3}$ | 26.84067 | $6.453 \cdot 10^{-6}$ | $9.074 \cdot 10^{-3}$ |
| $150^3$ | $6.6968 \cdot 10^{-3}$ | 7.59804 | $5.898 \cdot 10^{-5}$ | 0.1889 | $2.12490 \cdot 10^{-3}$ | 26.7625 | $6.024 \cdot 10^{-6}$ | $9.944 \cdot 10^{-3}$ |
| $200^3$ | $6.6976 \cdot 10^{-3}$ | 7.59686 | $5.751 \cdot 10^{-5}$ | 0.1810 | $2.12392 \cdot 10^{-3}$ | 26.7365 | $9.595 \cdot 10^{-6}$ | $2.260 \cdot 10^{-2}$ |
| $256^3$ | $6.6979 \cdot 10^{-3}$ | 7.59646 | $5.681 \cdot 10^{-5}$ | 0.1793 | $2.12389 \cdot 10^{-3}$ | 26.7231 | $1.093 \cdot 10^{-5}$ | $2.614 \cdot 10^{-2}$ |

Table 2. Estimation of the critical Grashof number for buoyancy convection in a laterally heated cube from time-dependent runs. References: [1] Janssen *et al.* (1993); [2] Janssen & Henkes, 1995; [3] Labrosse *et al.*, 1997; [4] De Gassowski *et al.*, 2006; [5] Janssen & Henkes, 1996; [6] Sheu *et al.*, 2011; [7] Soucasse *et al.*, 2014; [8] Jones & Briggs, 1989; [9] Gelfgat, 2007.

| Case  | $100^3$ | $150^3$ | $200^3$ | $256^3$ | [1,2], $120^3$ FV grid | [3] $61^3$ collocation points | [4] $60^3$ grid | [5] $120^3$ grid | [6] $81^3$ FD grid | [7] $80^3$ collocation points | [8] Experiment | [9] 2D problem $400^2$ FV grid |
|-------|---------|---------|---------|---------|------------------------|-------------------------------|-----------------|------------------|--------------------|-------------------------------|----------------|--------------------------------|
| CC-CC | $3.32 \cdot 10^6$ | $3.33 \cdot 10^6$ | $3.33 \cdot 10^6$ | $3.33 \cdot 10^6$ |  |  |  |  |  |  | $4.5 \cdot 10^6$ | $2.97 \cdot 10^6$ |
| CC-AA | $3.29 \cdot 10^6$ | $3.29 \cdot 10^6$ | $3.30 \cdot 10^6$ | $3.30 \cdot 10^6$ | $3.17 \cdot 10^6$ |  |  |  |  |  |  | $2.97 \cdot 10^6$ |
| AA-CC | $1.28 \cdot 10^8$ | $1.28 \cdot 10^8$ | $1.19 \cdot 10^8$ | $1.24 \cdot 10^8$ |  |  |  |  |  |  |  | $2.57 \cdot 10^8$ |
| AA-AA | $4.45 \cdot 10^7$ | $4.40 \cdot 10^7$ | $4.59 \cdot 10^7$ | $4.58 \cdot 10^7$ |  | $4.51 \cdot 10^7$ | $4.46 \cdot 10^7$ |  |  | $(4.4-4.9) \cdot 10^7$ |  |  |
| AA-AA | $6.62 \cdot 10^7$ | $6.95 \cdot 10^7$ | $7.12 \cdot 10^7$ | $7.2 \cdot 10^7$ |  |  |  |  |  |  |  |  |
| AA-AA | $2.33 \cdot 10^8$ | $2.76 \cdot 10^8$ | $2.80 \cdot 10^8$ | $2.80 \cdot 10^8$ |  |  |  | $(3.5-5.6) \cdot 10^8$ | $1.73 \cdot 10^8$ |  |  | $2.57 \cdot 10^8$ |

Table 3. Estimation of the oscillation frequency for buoyancy convection in a laterally heated cube from time-dependent runs. References are same as for Table 2.

| Case | $100^3$ | $150^3$ | $200^3$ | $256^3$ | [1,2], $120^3$ FV grid | [3] $61^3$ collocation points | [4] $60^3$ grid | [5] $120^3$ grid | [6] $81^3$ FD grid | [7] $80^3$ collocation points | [8] Experiment | [9] 2D problem $400^2$ FV grid |
|---|---|---|---|---|---|---|---|---|---|---|---|---|
| CC-CC | 0.281 | 0.282 | 0.282 | 0.282 | | | | | | | 0.26 | 0.254 |
| CC-AA | 0.263 | 0.264 | 0.264 | 0.264 | 0.266 | | | | | | | 0.254 |
| AA-CC | 0.157 | 0.00990 | 0.00990 | 0.00989 | | | | | | | | 0.05522 |
| AA-AA | 0.0174 | 0.0105 | 0.0105 | 0.0107 | | 0.009 | 0.0062 | | | 0.00795 | | |
| AA-AA | 0.106 | 0.104 | 0.105 | 0.105 | | | | | | | | |
| AA-AA | 0.0149 | 0.522 | 0.525 | 0.525 | | | | 0.532 | 0.3 | | | 0.05522 |

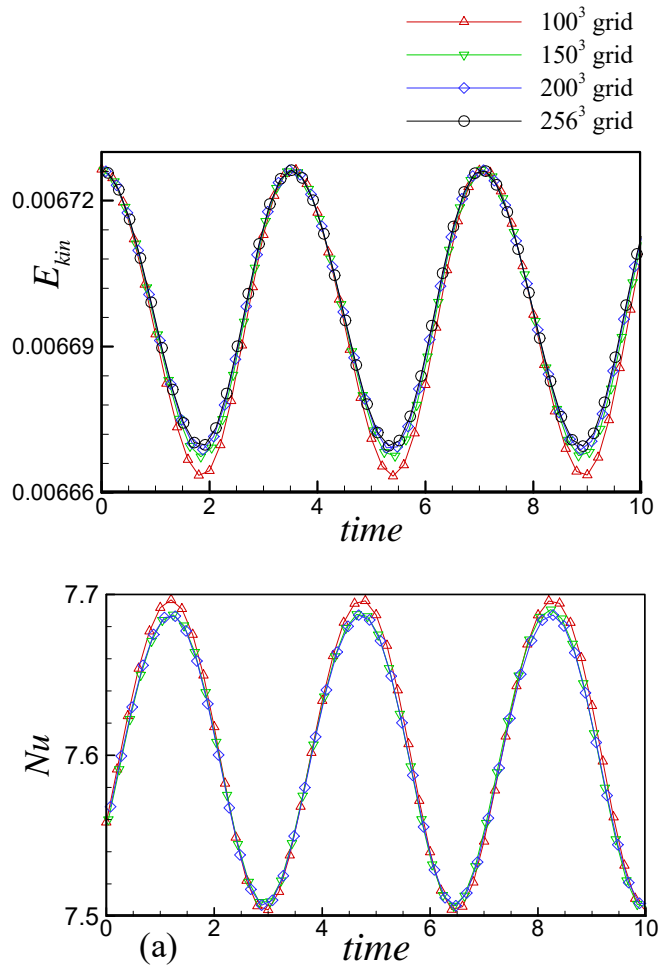
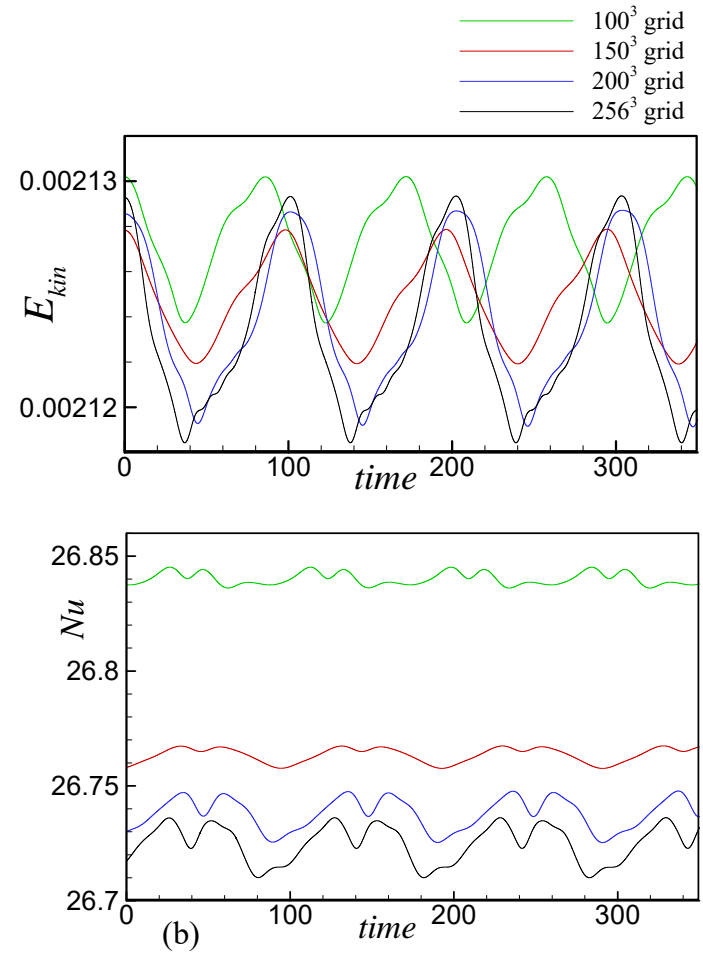

Figure 1. Flow oscillations at $Gr=3.4 \cdot 10^6$, CC – CC case (a), and $Gr=1.3 \cdot 10^8$, AA – CC case (right), at different finite volume grids.

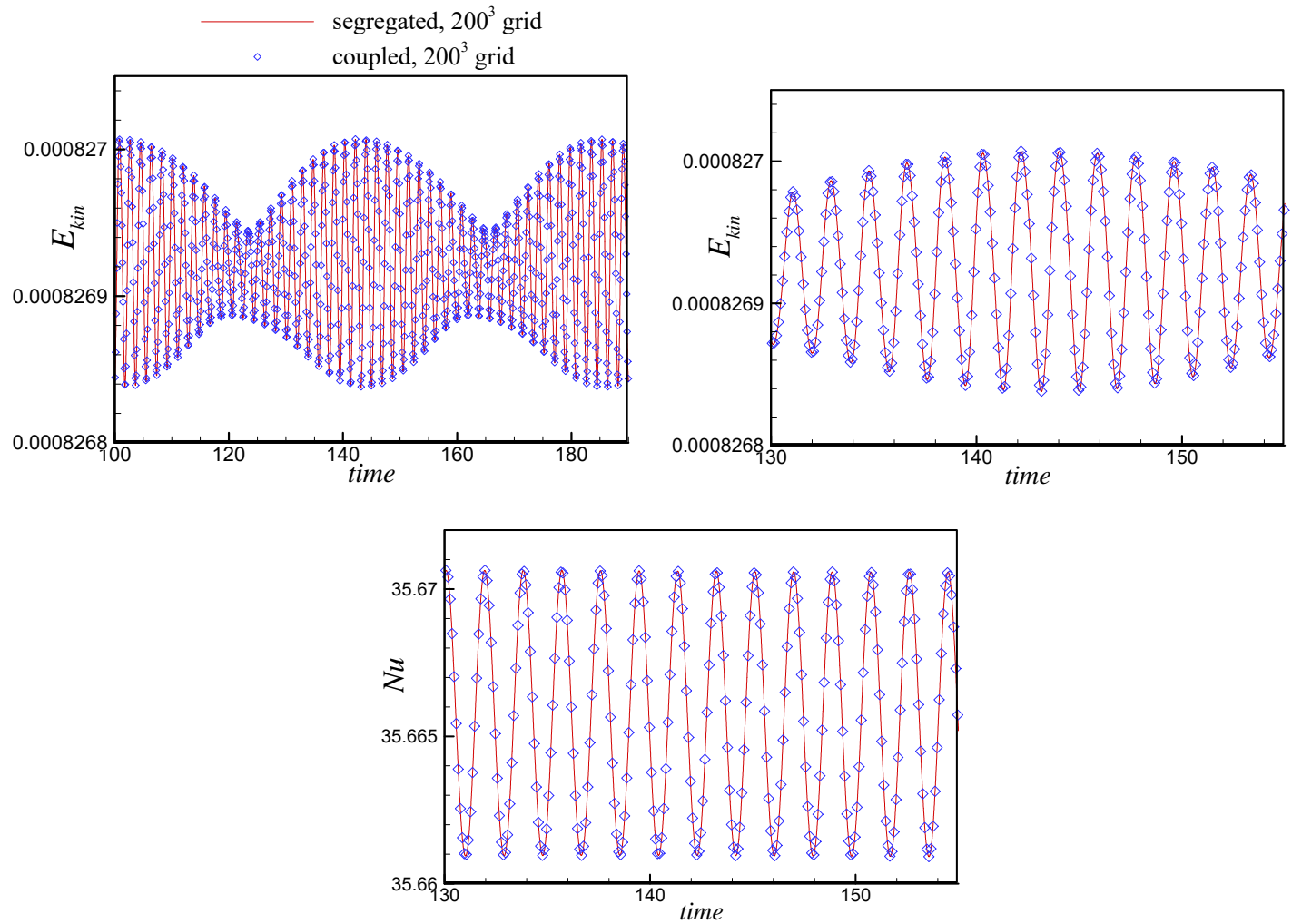

Figure 2. Comparison of time-dependent computations by pressure / velocity coupled and segregated solvers for two-frequency oscillatory regime calculated at Gr=3·10$^8$, AA – AA case, 200$^3$ grid.

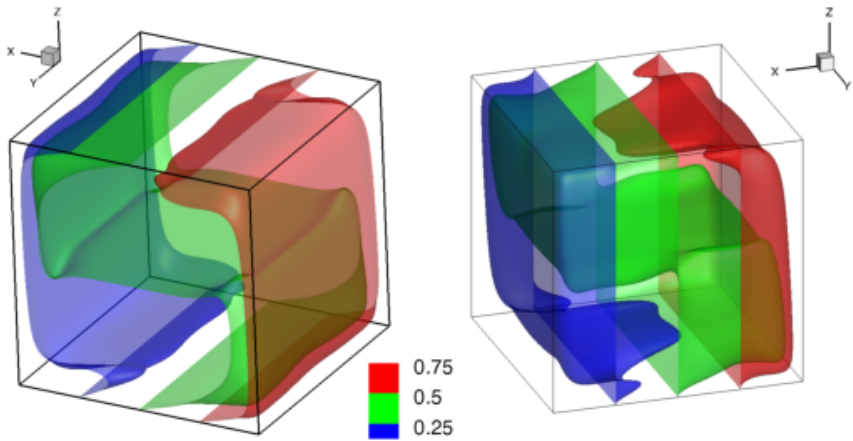
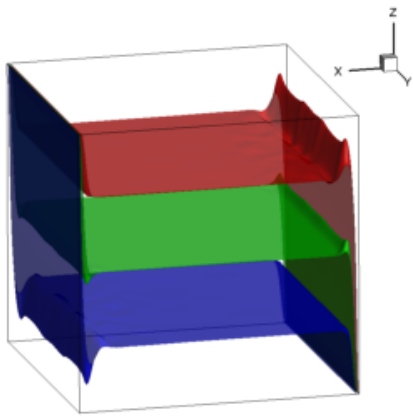
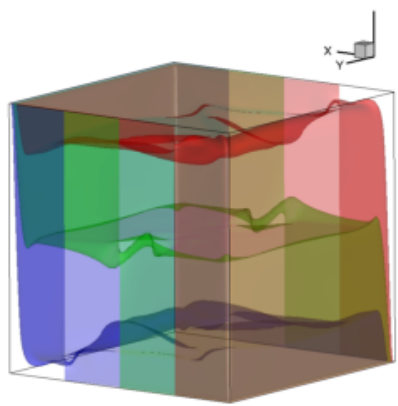

(a) CC - AA case, $Gr=2.9\times10^6$

(b) CC - CC case, $Gr=2.9\times10^6$

(c) AA - AA case, $Gr=2.2\times10^8$

(d) AA - CC case, $Gr=1.2\times10^8$

Figure 3. Temperature isosurfaces corresponding to slightly subcritical steady states.

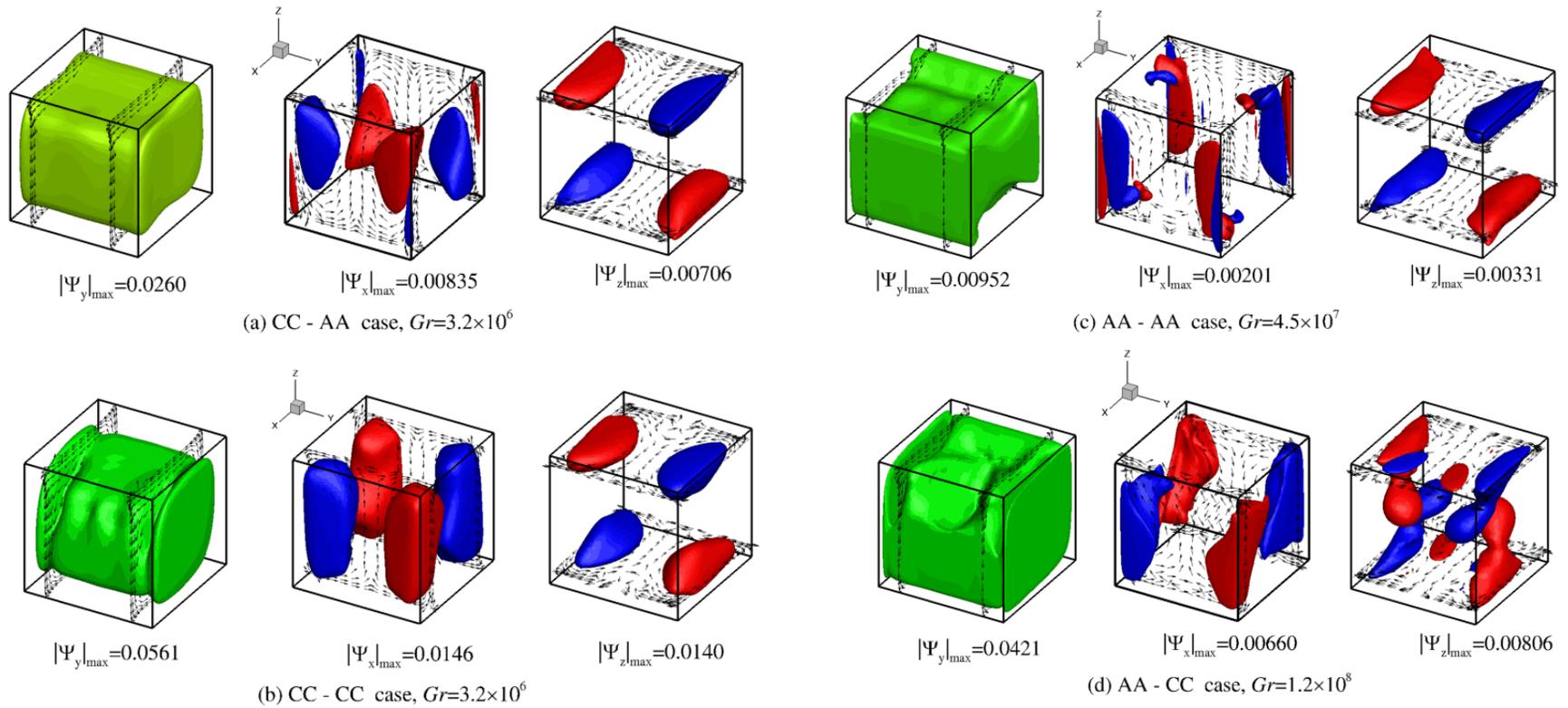

Figure 4. Visualization of 3D velocity fields corresponding to slightly subcritical steady states. Divergence free projections of velocity fields on the coordinate planes are shown by vectors. Isosurfaces of the velocity potentials, to which the vector fields are tangent, are shown by colors. In the frames (a), (b), and (c) the potential isosurfaces are plotted for the levels of 0.3 multiplied by the corresponding maximal absolute value. IN the frame (d) the factors are 0.5, 0.3, and 0.4 for $\Psi_y$, $\Psi_x$, and $\Psi_z$, respectively.

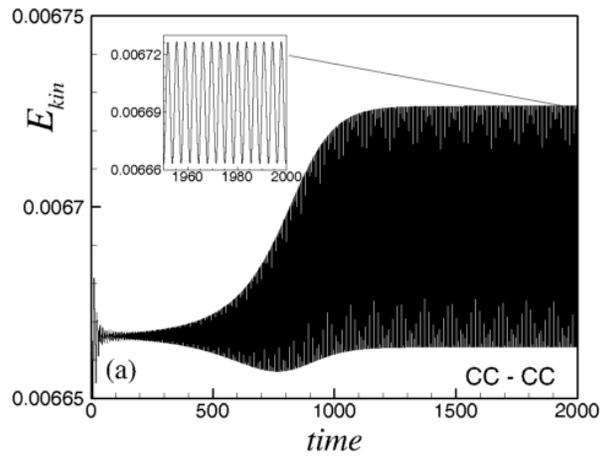
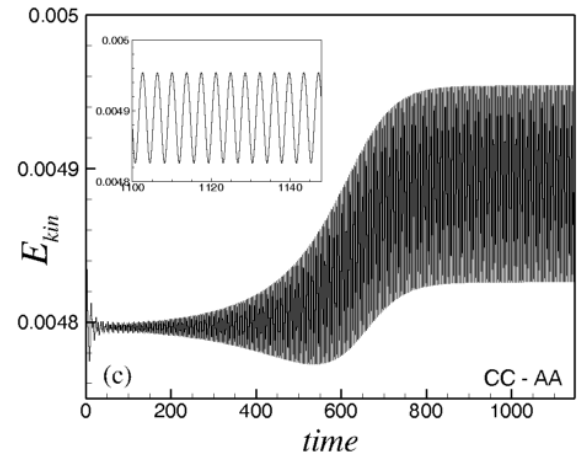
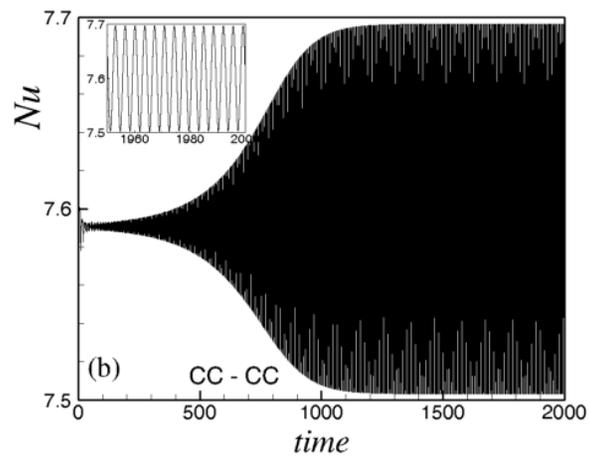
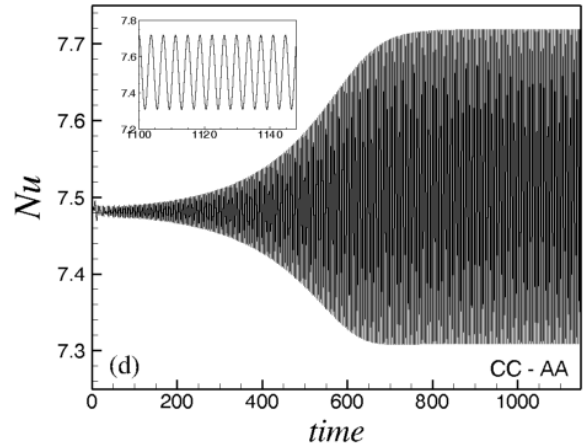

Figure 5. Time evolution of the total kinetic energy and the Nusselt number from $Gr=3.4 \cdot 10^6$ to $Gr=3.5 \cdot 10^6$. (a) and (b) CC – CC case; (c) and (d) CC – AA case. Calculation on the $100^3$ grid.

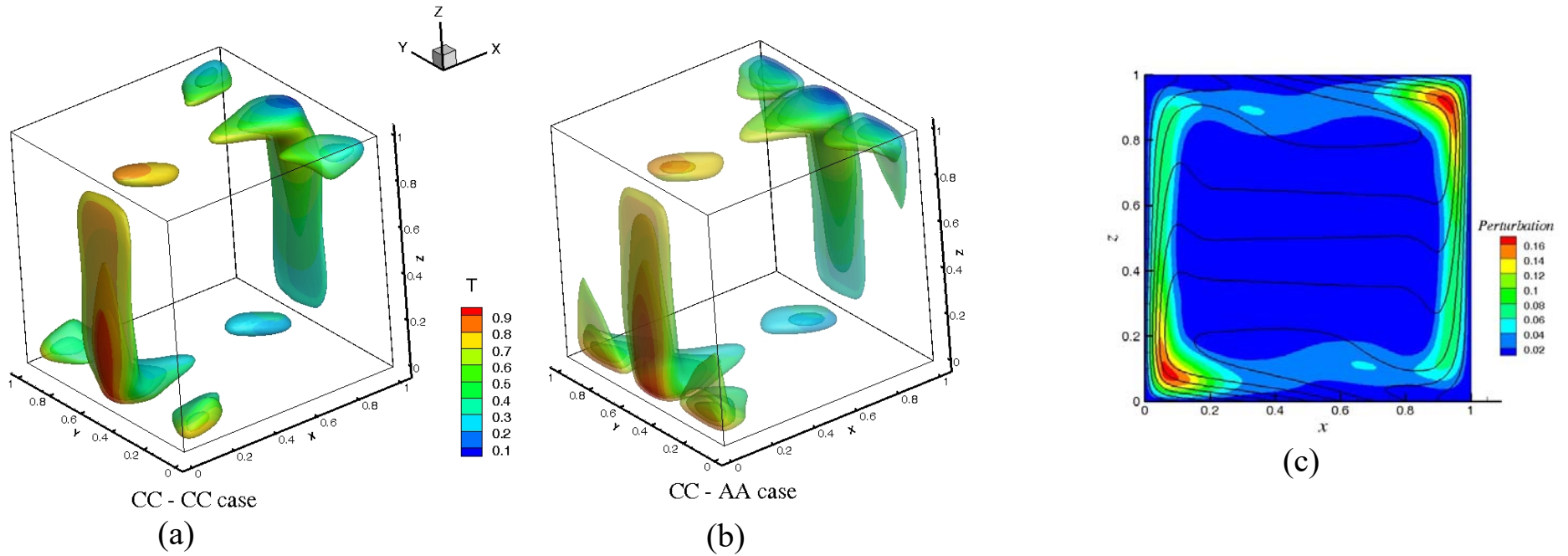

Figure 6. Amplitude of temperature oscillations at $Gr=3.4\cdot10^6$ in CC – CC (a) and CC – AA (b) cases. Colors of the isosurfaces correspond to the average temperature values. (c) Isotherms (black lines) and amplitude of the most unstable temperature perturbation (colors) in the 2D case with perfectly conducting horizontal boundaries. Calculation on the $100^3$ grid. Maximal values of the amplitude and the levels plotted are, respectively, 0.0769, 0.01, 0.02, 0.03 in the frame (a) and 0.114, 0.015, 0.025, 0.035 in the frame (b).

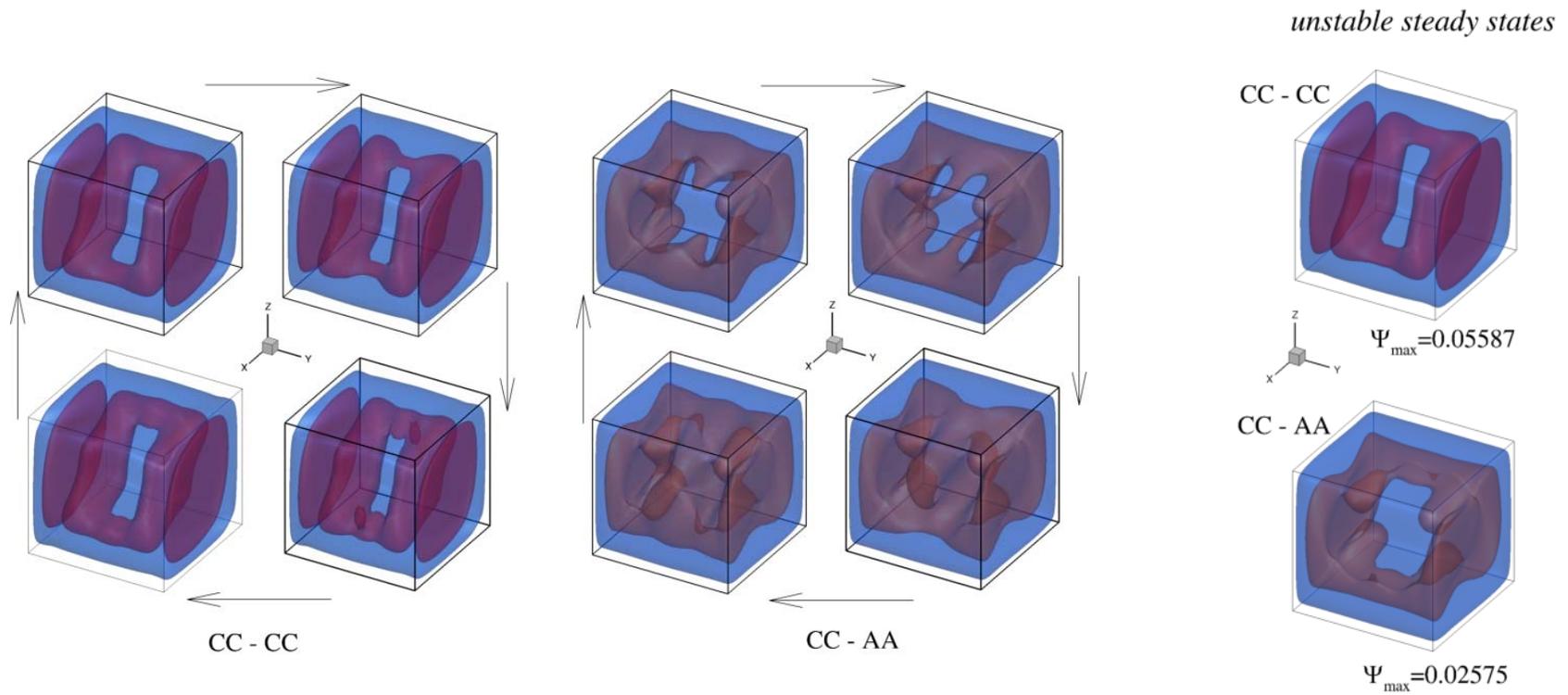

Figure 7. Snapshots of isosurfaces of vector potential $\Psi_y$ corresponding to the main convective circulation compared to the patterns of unstable steady state flows. $Gr=3.4\times10^6$. The plotted levels are 0.00335 and 0.0223 for the CC – CC case and 0.0015 and 0.0195 for the CC – AA case. Calculation on the $100^3$ grid. Animation1. Animation2.

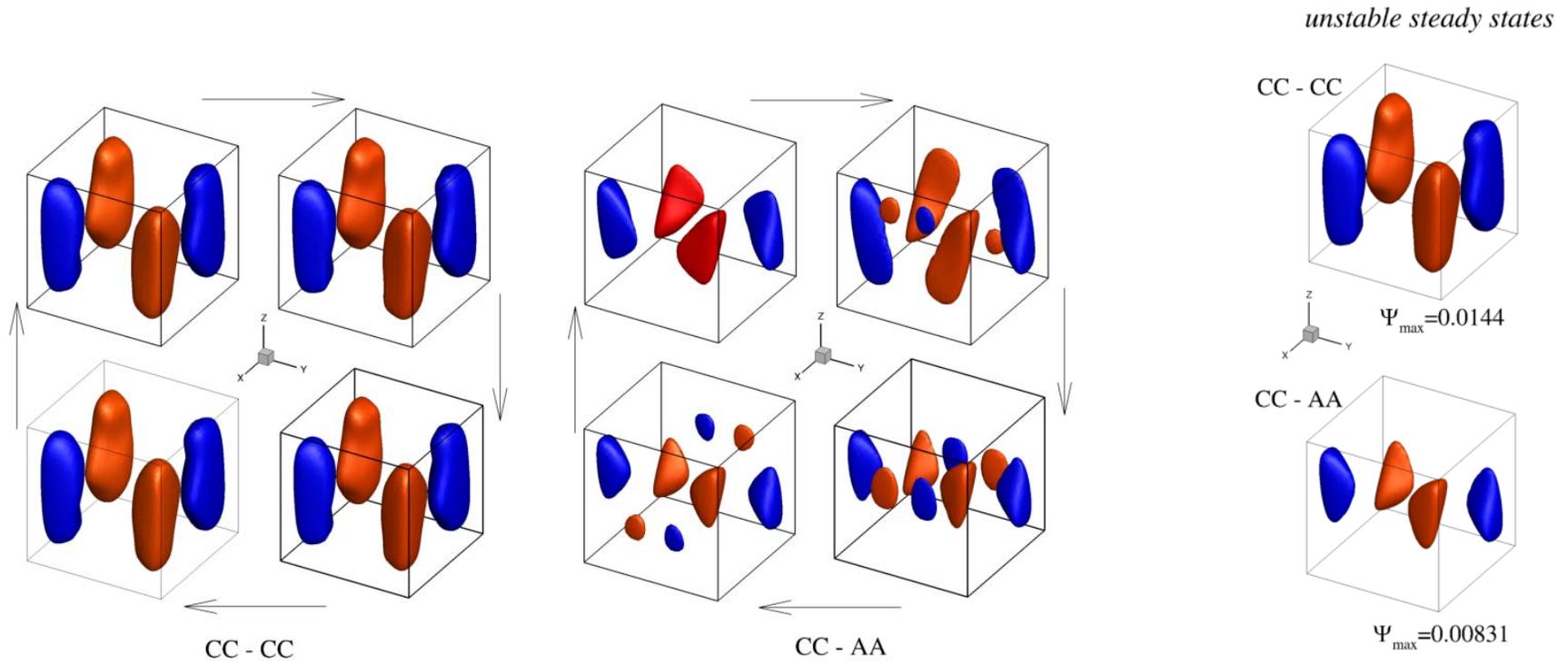

Figure 8. Snapshots of isosurfaces of vector potential $\Psi_x$ compared to the patterns of unstable steady state flows. $Gr=3.4\times10^6$. CC – CC case (left frames), and CC – AA case (right frames). The plotted levels are $\pm0.0073$ for the CC – CC case and $\pm0.004$ the CC – AA case. Calculation on the $100^3$ grid. Animation1. Animation2.

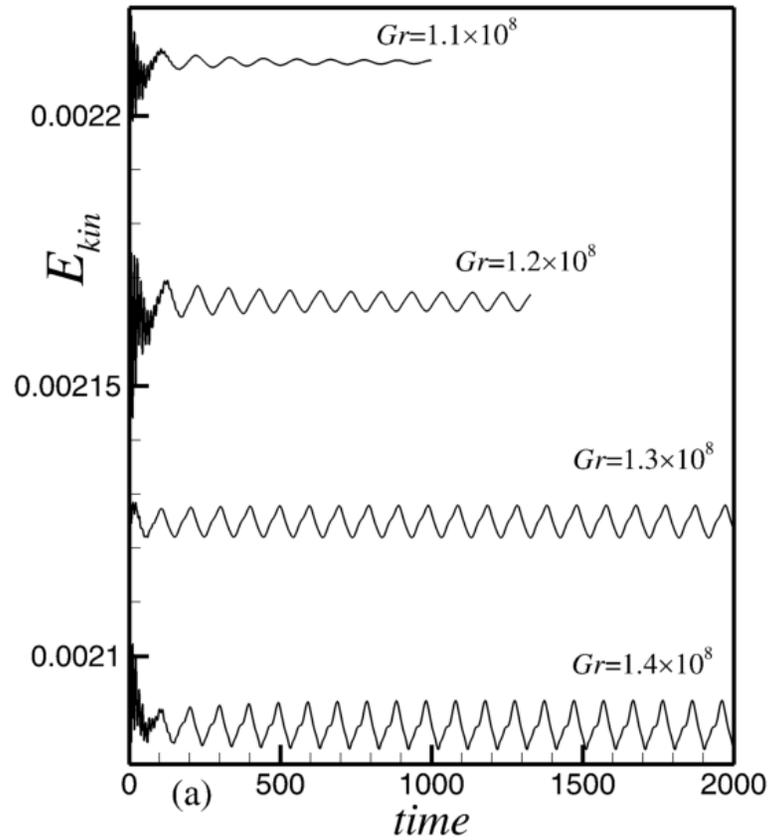 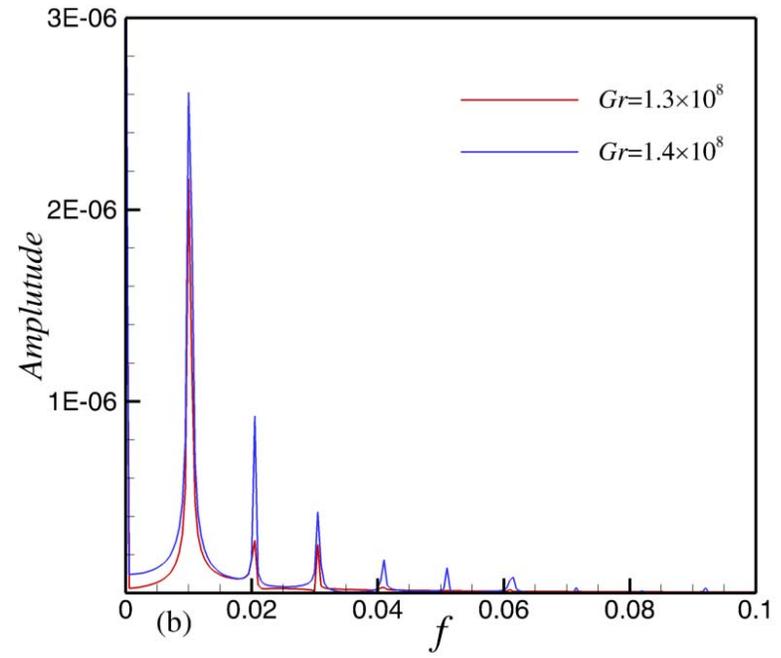

Figure 9. Time evolution of the total kinetic energy at $Gr=1.1 \cdot 10^8$ to $Gr=1.4 \cdot 10^8$ for the AA – CC case (a) and the frequency spectrum for the two largest values of the Grashof number. Calculation on the $150^3$ grid.

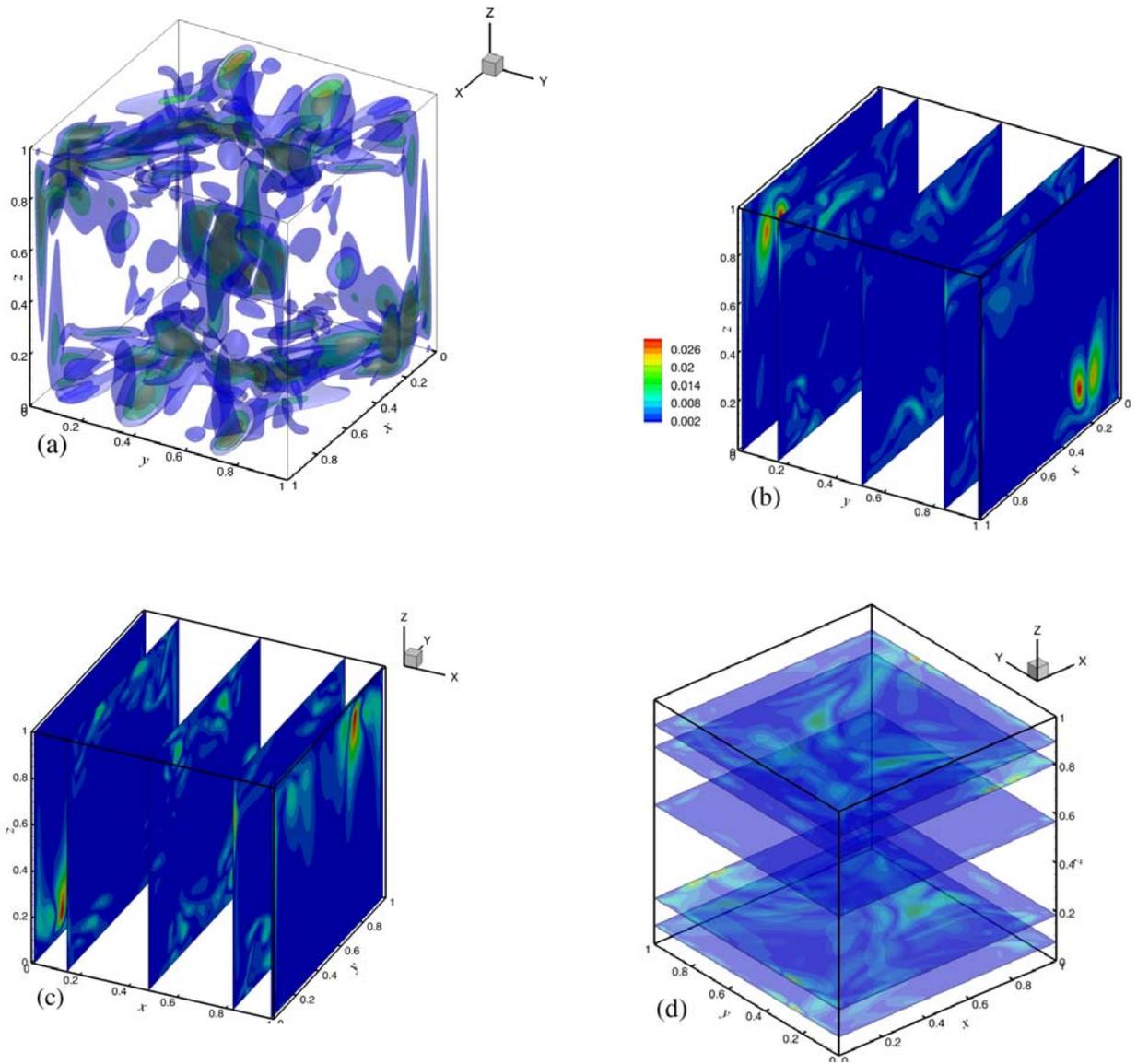

Figure 10. (a) Isosurfaces of amplitude of temperature oscillations at $Gr=1.3\cdot10^8$ in AA – CC case. The maximal value and levels plotted are 0.0302, 0.004, 0.007, 0.01. (b), (c), and (d) Amplitude isolines in characteristic $y$-, $x$-, and $z$- cross-sections respectively. Calculation on the $200^3$ grid. Animation 3.

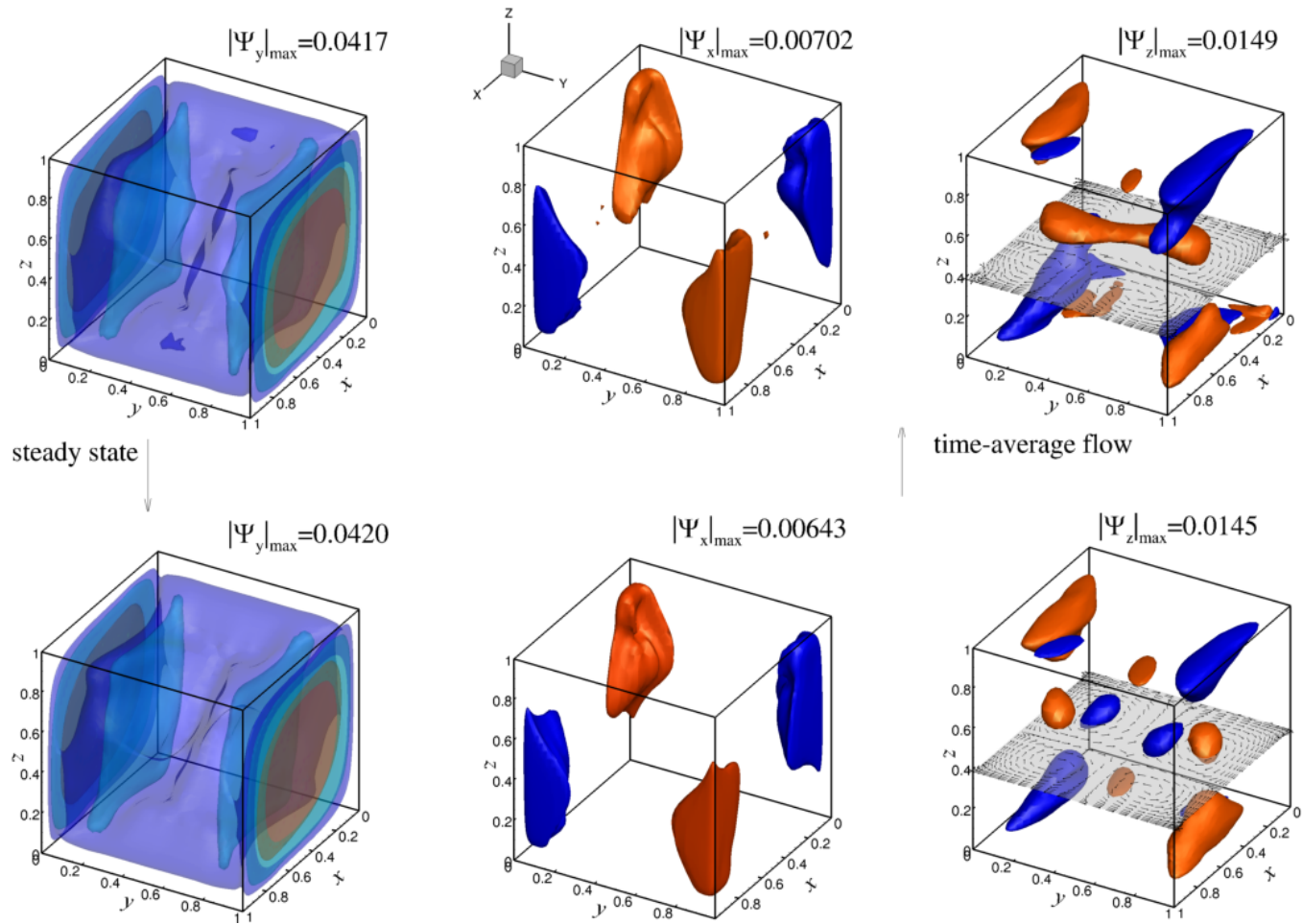

Figure 11. Comparison of vector potentials of the time-averaged oscillatory flow states at $Gr=1.3 \cdot 10^8$ in AA – CC (upper frames) with those of the unstable steady state at the same Grashof number (lower frames). Isosurface are shown at $\Psi_y =$ 0.006, 0.01 and 0.02, $\Psi_x = \pm 0.004$, and $\Psi_z = \pm 0.004$. Calculation on the $200^3$ grid.

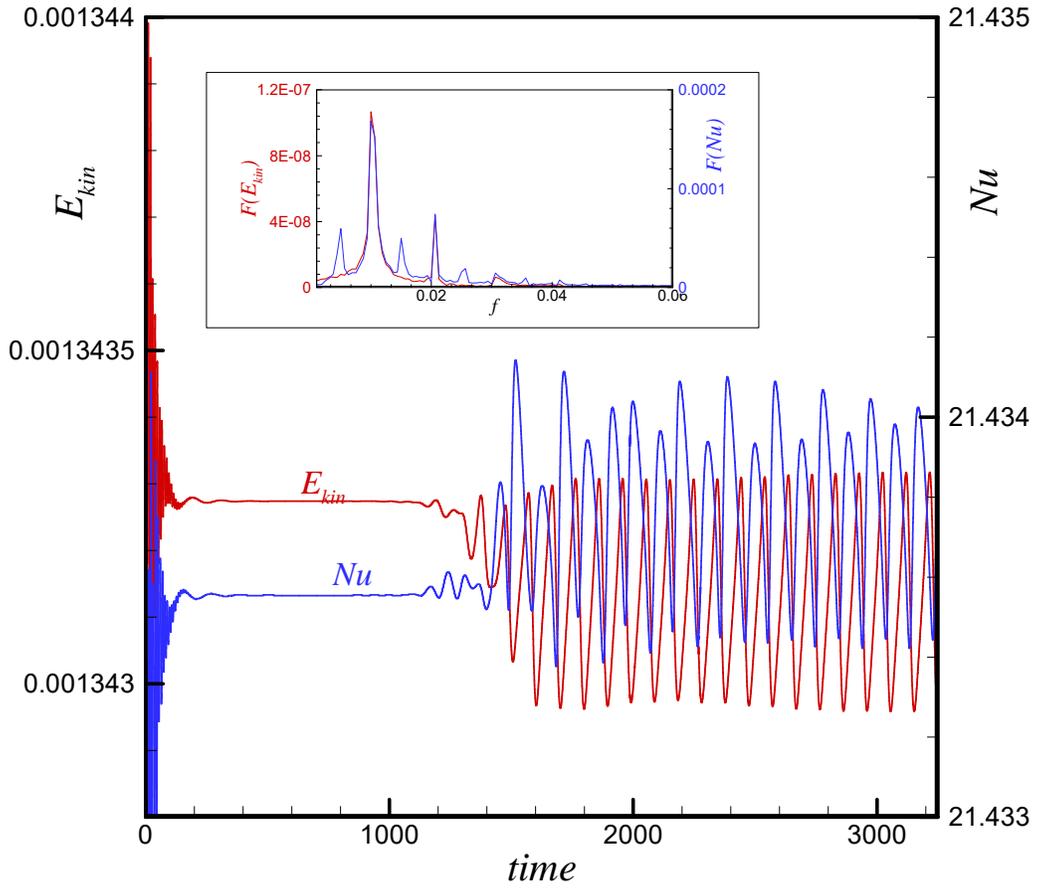

Figure 12. Time evolution of the total kinetic energy and Nusselt number at $Gr=4.6\cdot10^7$ starting from steady state at $Gr=4.55\cdot10^7$ in the AA – AA case. Calculation on the $200^3$ grid.

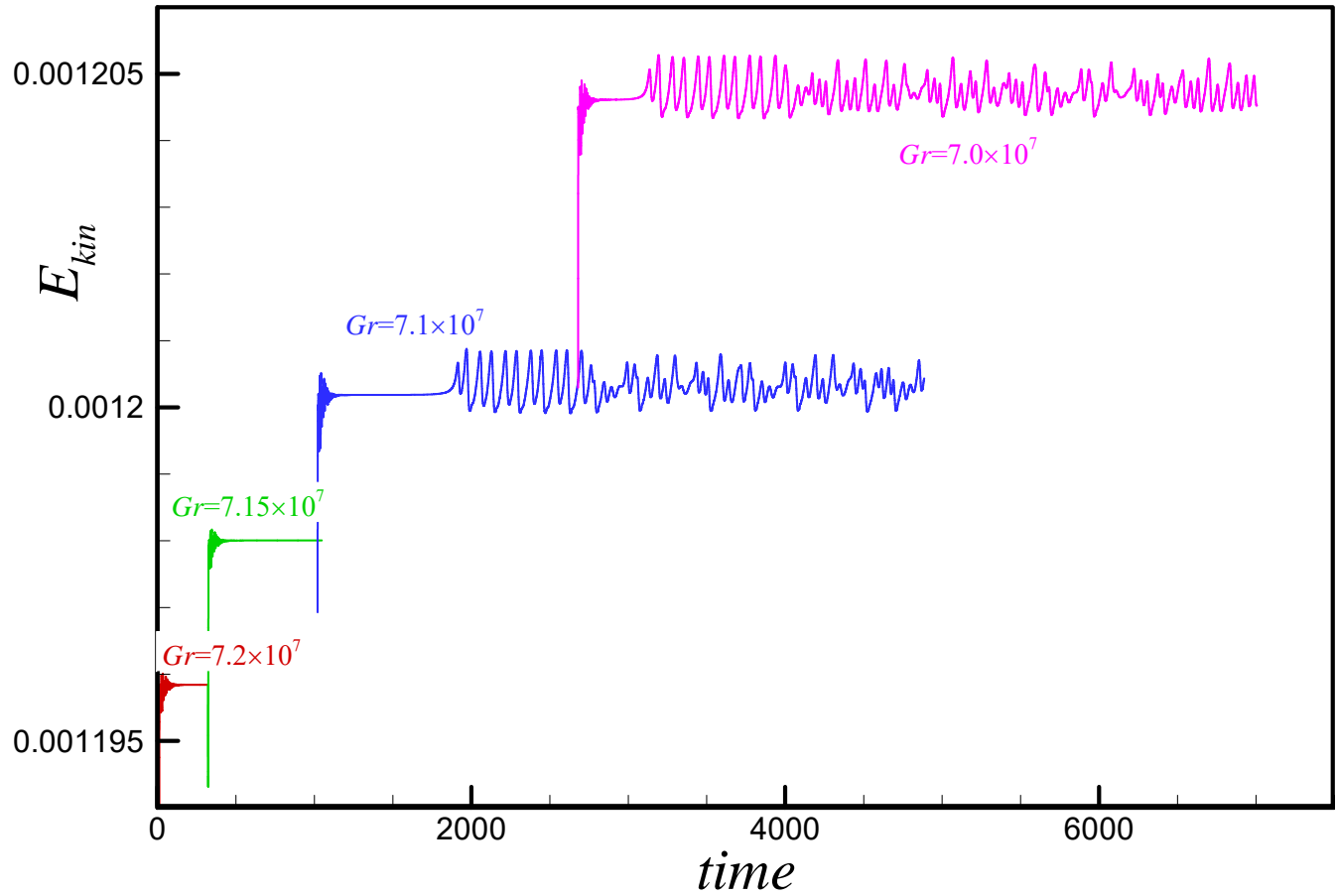

Figure 13. Time evolution of the total kinetic energy from $Gr=7.2 \cdot 10^7$ up to $Gr=7.0 \cdot 10^7$ starting from steady state at $Gr=7.25 \cdot 10^7$ in the AA – AA case. Calculation on the $200^3$ grid.

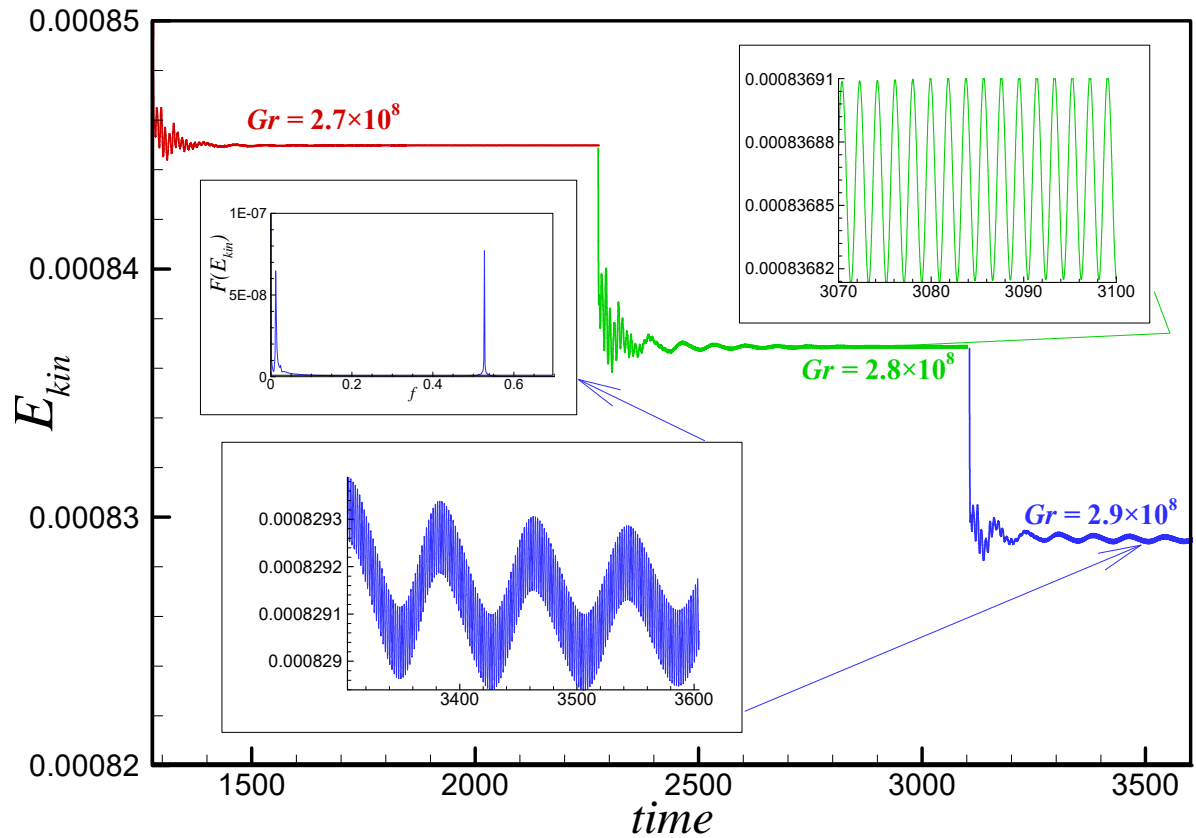

Figure 14. Time evolution of the total kinetic energy from $Gr=2.6·10^8$ to $Gr=2.7·10^8$ (red line), then to $Gr=2.8·10^8$ (green line), and then to $Gr=2.9·10^8$ (blue line). AA – AA case. Calculation on the $200^3$ grid.

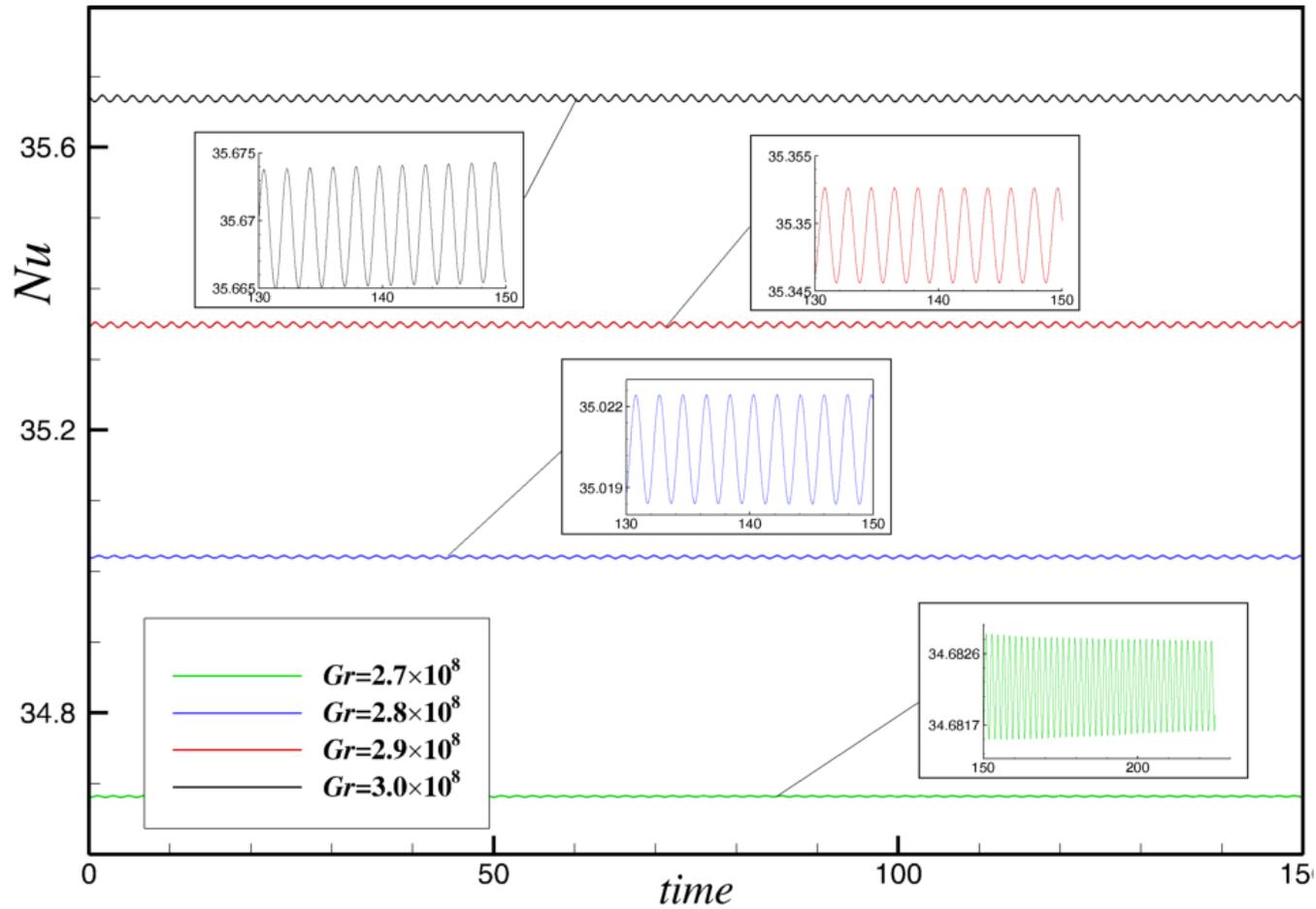

Figure 15. Large frequency / small amplitude oscillations of the Nusselt number computed for gradually increased Grashof number in AA – AA case. Calculation on the $200^3$ grid.

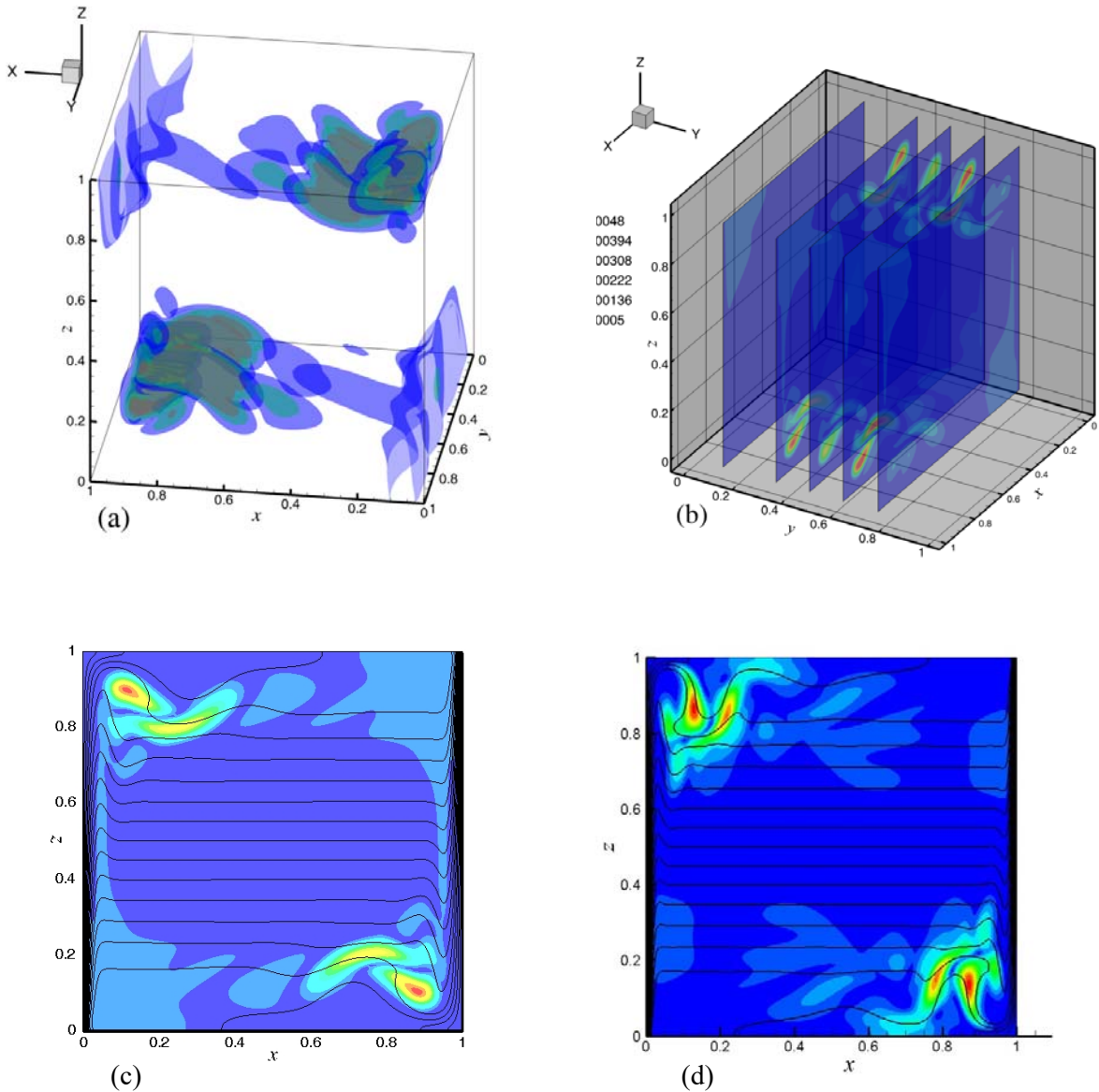

Figure 16. (a) Isosurfaces of amplitude of temperature oscillations at $Gr=4.7\cdot 10^7$ in AA – AA case. (b) Amplitude of the temperature oscillations in several $y=const$ planes of the frame (a). (c) Isotherms (black lines) and amplitude of the temperature oscillations (colors) in $y=0.35$ plane of the frame (a). (d) Isotherms (black lines) and amplitude of the temperature oscillations (colors) of the most unstable perturbation in the 2D AA case. Calculation on the $150^3$ grid. Animation 4.

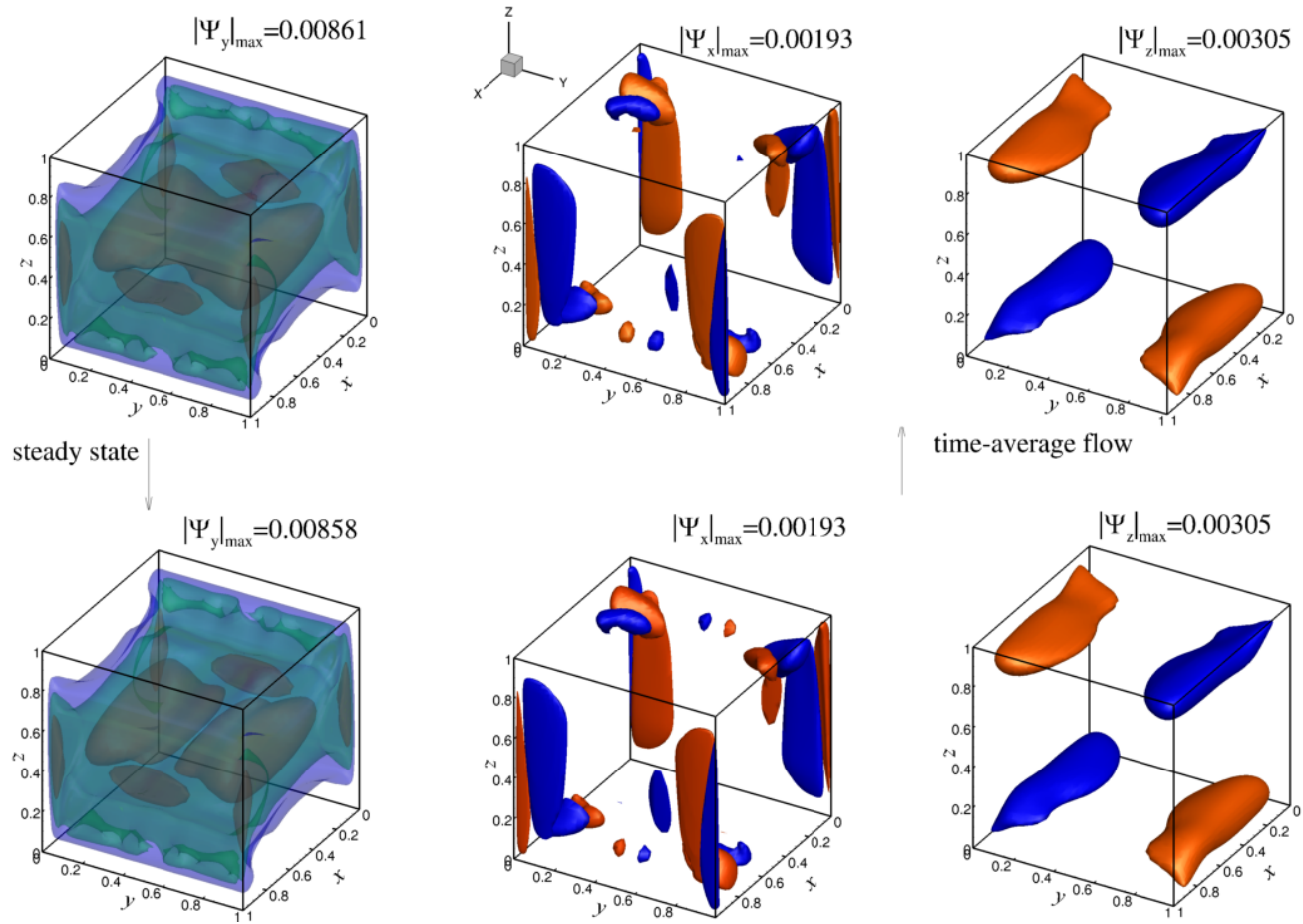

Figure 17. Comparison of vector potentials of the time-averaged oscillatory flow states at $Gr=7\cdot 10^7$ in AA – AA (upper frames) with those of the unstable steady state at the same Grashof number (lower frames). Isosurface are shown at $\Psi_y =$ 0.0045, 0.006 and 0.0075, $\Psi_x = \pm 0.0006$, and $\Psi_z = \pm 0.001$. Calculation on the $200^3$ grid.

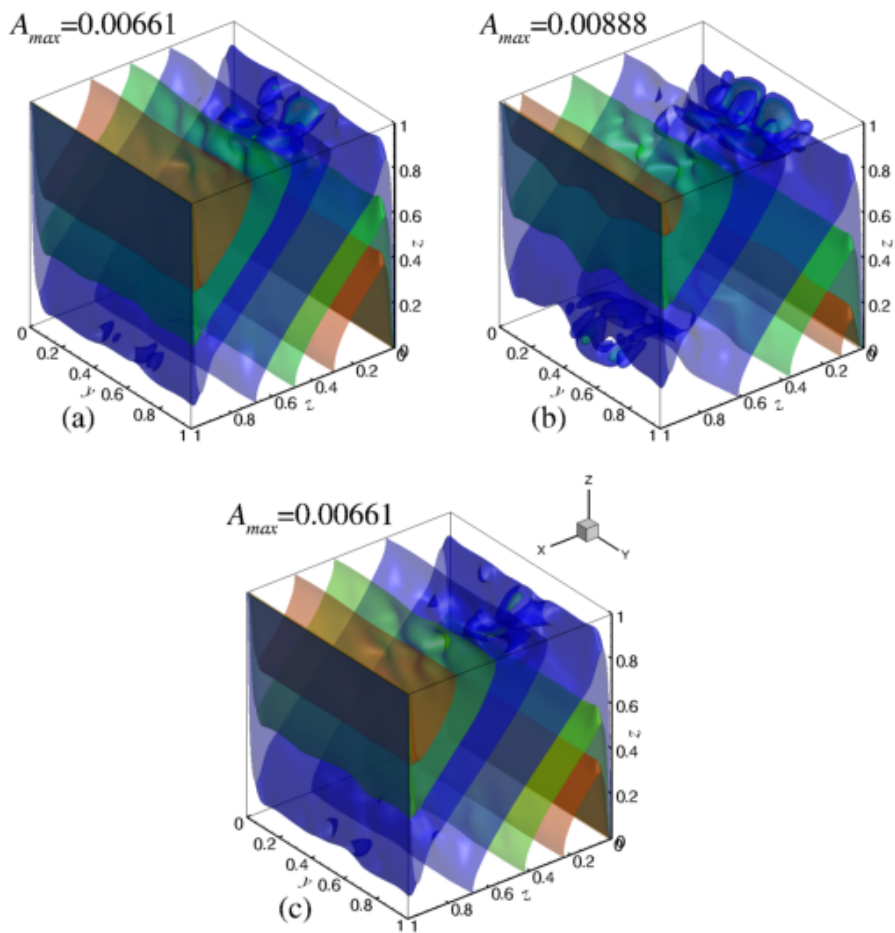

Figure 18. Amplitudes of three-frequent temperature oscillations at $Gr=7\cdot10^7$, AA – AA case. The values of dimensionless frequencies are 0.00610, 0.0122, and 0.0183 for the frames (a), (b), and (c), respectively. Calculation on the $200^3$ grid.

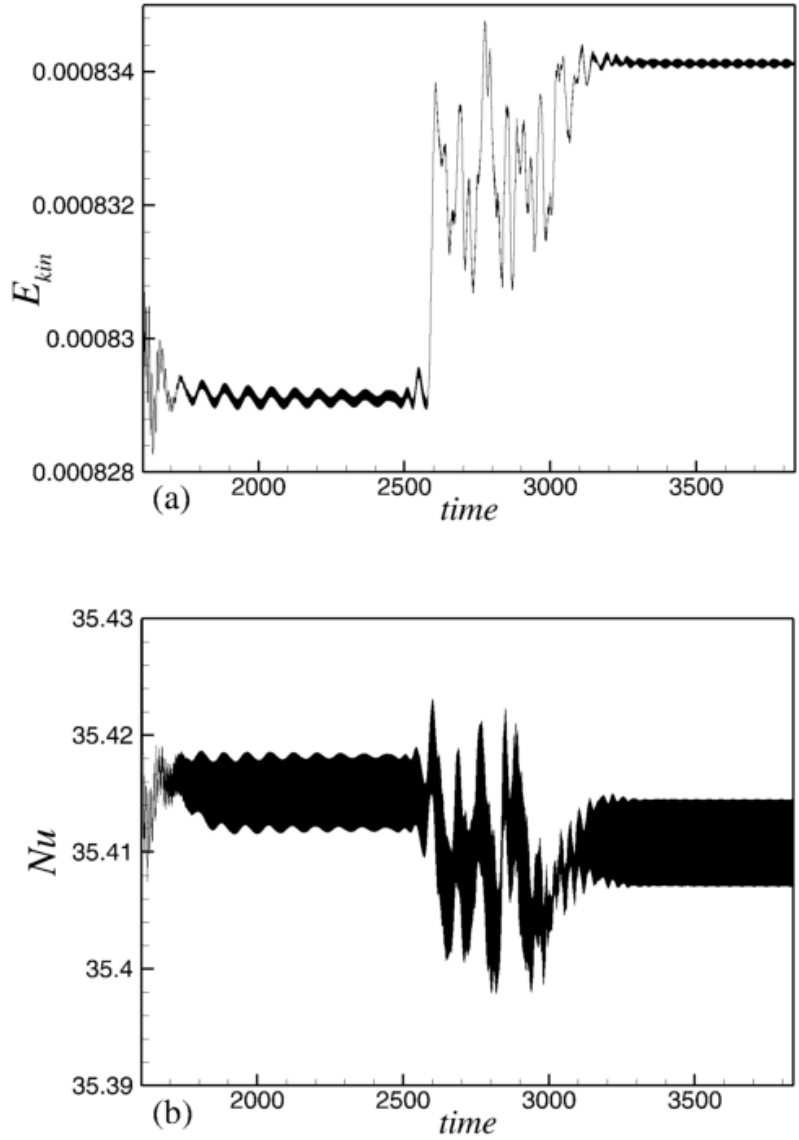

Figure 19. Time evolution of the total kinetic energy (a) and Nusselt number (b) at $Gr=2.9 \cdot 10^8$. AA – AA case. Calculation on the $150^3$ grid.

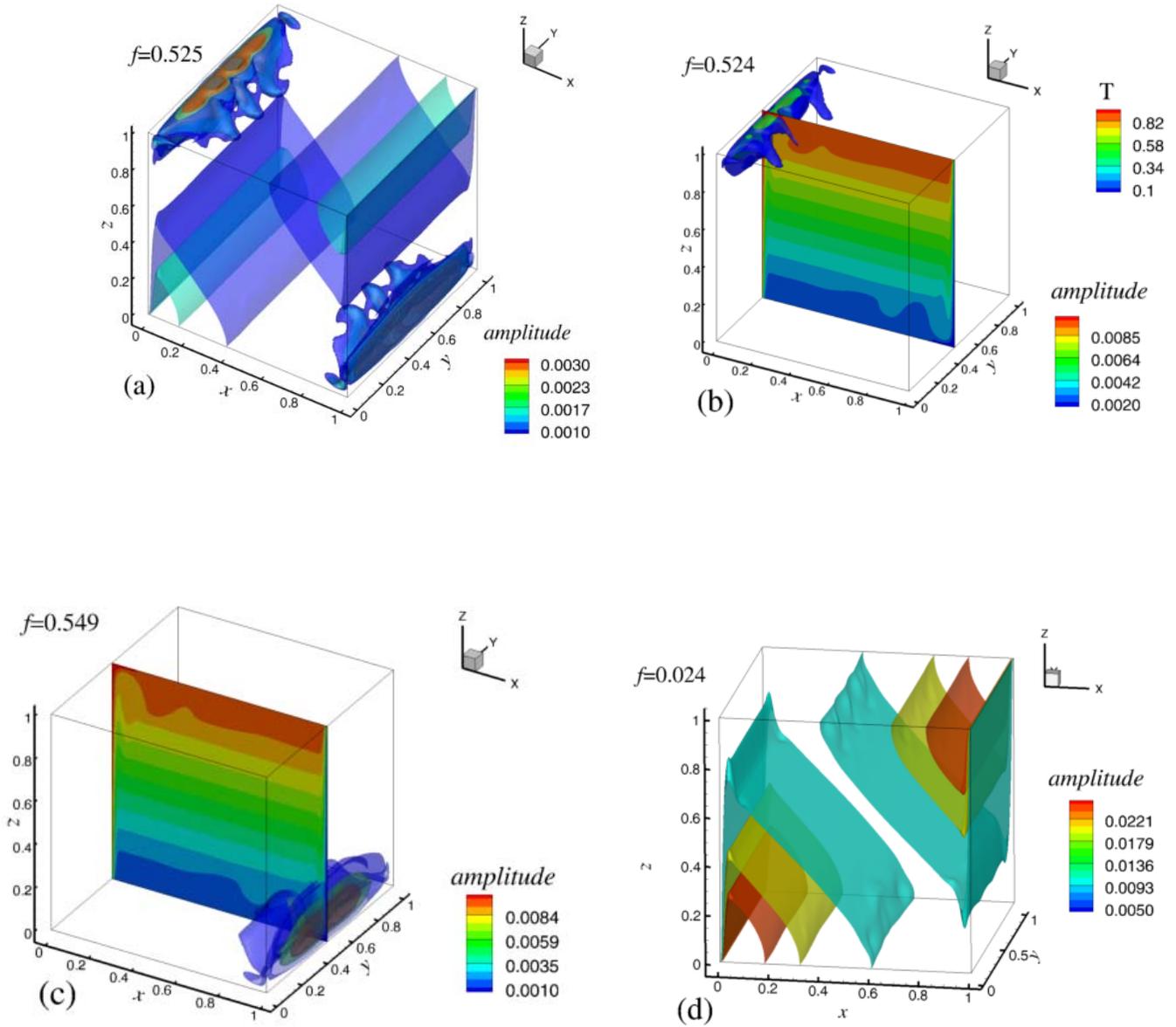

Figure 20. Amplitudes of temperature oscillations at $Gr=2.9 \cdot 10^8$, AA – AA case. (a) amplitude of the single frequency regime shown in Fig. 15. (b) – (d) amplitudes of the three main frequencies of the second oscillatory regime shown in Figs. 2b and 17. Calculation on the $150^3$ grid.

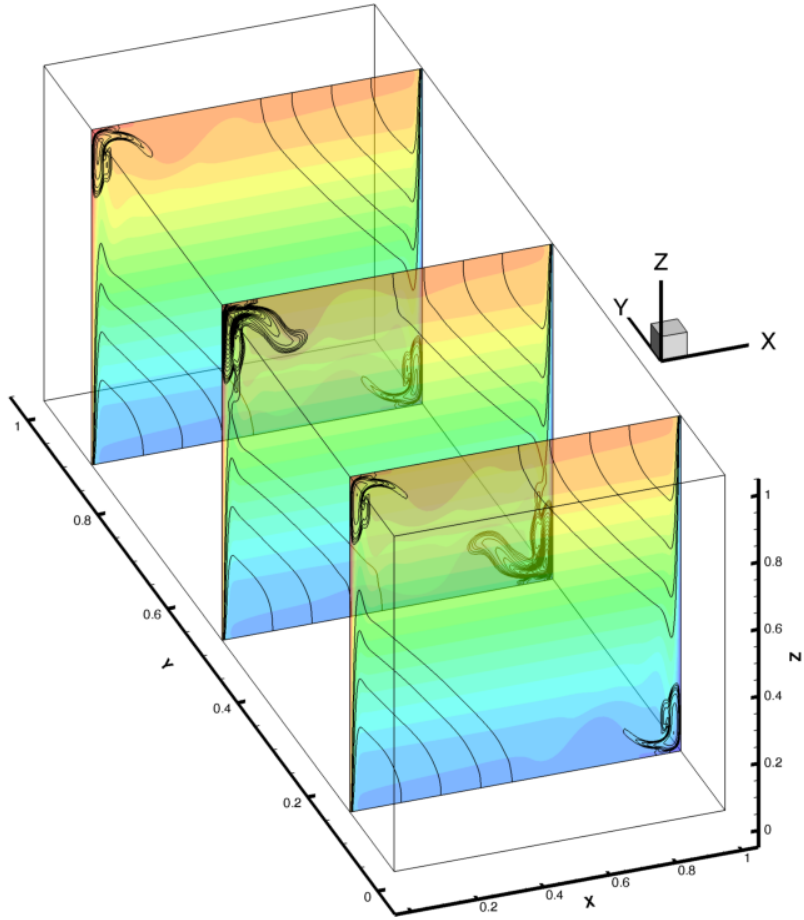

Figure 21. Time-averaged isotherms (colors) and amplitude of the single frequency oscillatory regime at $Gr=2.9 \cdot 10^8$, AA – AA case. Calculation on the $150^3$ grid.